\documentclass[prb,twocolumn,showpacs,superscriptaddress,floatfix, nofootinbib]{revtex4-2}
\usepackage{amssymb, amsfonts, amsmath, mathrsfs, amsbsy, dsfont}
\usepackage{graphicx}
\usepackage[bookmarks=true, colorlinks=true, linkcolor=blue, urlcolor=blue, citecolor=blue, bookmarks=true, hyperindex=true]{hyperref}

\newcommand{\be}{\begin{equation}}
\newcommand{\ee}{\end{equation}}

\newcommand{\beq}{\begin{eqnarray}}
\newcommand{\eeq}{\end{eqnarray}}

\def\H1{\widehat{H}_1}

\renewcommand{\i}{\ensuremath{\mathrm{i}}}

\renewcommand{\d}{\ensuremath{\mathrm{d}}}

\newcommand{\braket}[1]{\left\langle #1 \right\rangle}
\newcommand{\ket}[1]{\left| #1 \right\rangle}
\newcommand{\bra}[1]{\left\langle #1 \right|}

\renewcommand{\L}{\pmb{\mathcal{L}}}

\renewcommand{\b}{\pmb{b}}
\renewcommand{\d}{\pmb{d}}
\newcommand{\rphi}{\overrightarrow{\phi}}
\newcommand{\lphi}{\overleftarrow{\phi}}

\newcommand{\rpsi}{\overrightarrow{\psi}}
\newcommand{\lpsi}{\overleftarrow{\psi}}
\newcommand{\rbpsi}{\overrightarrow{\pmb{\psi}}}
\newcommand{\lbpsi}{\overleftarrow{\pmb{\psi}}}

\DeclareMathOperator{\tr}{tr}

\DeclareMathOperator{\im}{Im}

\newcommand{\bbraket}[1]{ \braket{\!\braket{#1} \!} }
\newcommand{\bbraketvac}[1]{\bbraket{\mathds{1}|#1|\rho_0}}

\begin{document}

\title{Conformal symmetry in quasi-free Markovian open quantum systems}

\author{Anatoliy\,I. Lotkov}
\affiliation{Department of Physics, University of Basel, Klingelbergstrasse 82, CH-4056 Basel, Switzerland}
\affiliation{Russian Quantum Center, Skolkovo, Moscow 121205, Russia}

\author{Denis\,V.~Kurlov}
\affiliation{Department of Physics, University of Basel, Klingelbergstrasse 82, CH-4056 Basel, Switzerland}
\affiliation{Russian Quantum Center, Skolkovo, Moscow 121205, Russia}
\affiliation{National University of Science and Technology ``MISIS'', Moscow 119049, Russia}

\author{Aleksey K. Fedorov}
\affiliation{Russian Quantum Center, Skolkovo, Moscow 121205, Russia}
\affiliation{National University of Science and Technology ``MISIS'', Moscow 119049, Russia}

\author{Nikita A. Nemkov}
\affiliation{Russian Quantum Center, Skolkovo, Moscow 121205, Russia}
\affiliation{National University of Science and Technology ``MISIS'', Moscow 119049, Russia}

\author{Vladimir Gritsev}
\affiliation{Institute for Theoretical Physics Amsterdam, University of Amsterdam, P.O. Box 94485, 1090 GL Amsterdam, The Netherlands}
\affiliation{Russian Quantum Center, Skolkovo, Moscow 121205, Russia}

\begin{abstract}
Conformal symmetry governs the behavior of closed systems near second-order phase transitions, and is expected to emerge in open systems going through dissipative phase transitions. We propose a framework allowing for a manifest description of conformal symmetry in open Markovian systems. The key difference from the closed case is that both conformal algebra and the algebra of local fields are realized on the space of superoperators. We illustrate the framework by a series of examples featuring systems with quadratic Hamiltonians and linear jump operators, where the Liouvillian dynamics can be efficiently analyzed using the formalism of third quantization.
We expect that our framework can be extended to interacting systems using an appropriate generalization of the conformal bootstrap.
\end{abstract}

\maketitle

\section{Introduction}

Understanding different phases of matter and phase transitions is one of the central themes in contemporary physics. In classical systems phase transitions are driven by thermal fluctuations, whereas in quantum systems they can also occur at zero temperature due to quantum fluctuations \cite{Sachdev2011}. Phase transitions in equilibrium quantum systems are associated with a non-analytic behavior of observables in the ground state and with the closure of the Hamiltonian gap.
The gap closure signifies the absence of dimensional parameters, which results in the scale invariance, unless the system exhibits a conformal anomaly \cite{DiFrancesco1997}. 
In physically relevant two-dimensional systems, the conformal symmetry follows directly from the scale invariance ~\cite{Polchinski1988, Nakayama2014}. Consequently, conformal field theories (CFTs) can describe two-dimensional classical and one-dimensional quantum systems at the critical point~\cite{DiFrancesco1997}.

Dissipative systems display non-equilibrium phases of matter with no equilibrium counterparts~\cite{Vojta2003}. Notable examples include the dissipative phase transition described by the Kardar-Parisi-Zhang equation~\cite{Kardar1986,Krug1997,HalpinHealy1995}, the measurement-induced phase transitions~\cite{Aharonov2000_PhysRevA.62.062311, Fisher2019_PhysRevB.100.134306,Skinner2019_PhysRevX.9.031009,Iaconis2020}, and dissipative time crystals~\cite{Gong2018, Buca219,  Muniz2020,  Roberts2020, Keler2020, Ippoliti2021, Bakker2022}. 
Dissipative phase transitions have also been experimentally observed in Rydberg atom  systems~\cite{Ding_PhysRevX.10.021023,Ding_2022} and may serve as a resource for precision 
measurements~\cite{Zanardi2008}.

Similarly to the case of isolated systems, phase transitions in open systems are characterized by a non-analytic behavior of observables in the steady state and by the closure of the Liouvillian gap \cite{Minganti2018,Henkel2010}. 
Various approaches to dissipative phase transition we examined~\cite{Heyl2018, Vojta2003,Cubitt2015}. 
Scale invariance and full conformal symmetry are expected to emerge in open systems as well, potentially associated with non-unitary or non-diagonalizable field theories \cite{Cardy2013,Creutzig2013}. Dissipative phase transitions are also closely related to fixed points of the non-equilibrium renormalization group~\cite{Berges2008, Berges2009,Berges2012, Nowak2012,Schole2012,Karl2013}. 
The investigation of conformal symmetry in dissipative systems has gained a lot of interest in recent years~\cite{Cristofano2004, Nakamura2012, Dutta2015, Chang2020}. However, a comprehensive description remains elusive thus far.

A state of an open quantum system, i.e. a system interacting with an environment, is described by a density matrix. 
Under the assumption of Markovianity, the evolution of the density matrix is described by the Lindblad equation \cite{Lindblad1976, Gorini1976}
\begin{align}
    \partial_t\rho=\L \rho:=-i[H,\rho]+\sum_{k} \left(L_k \rho L_k^\dagger - \frac12\left\lbrace L_k^\dagger L_k,\rho\right\rbrace\right), \label{Lindblad equation}
\end{align}
where $H$ is the Hamiltonian of the system and the jump operators $L_k$ can be interpreted as elementary interactions with the environment. The Liouvillian superoperator $\L$ is the generator of time evolution and can thus be viewed as a generalization of the Hamiltonian to open systems. Hereinafter we refer to linear maps on density matrices as superoperators and denote them with bold symbols. 

Importantly, the Liouvillian is not in general Hermitian with respect to the Hilbert-Schmidt scalar product 
\begin{align}
\bbraket{\rho_1 | \rho_2}:=\operatorname{Tr} \rho_1^\dagger \rho_2   \ , \label{Hilbert-Schmidt}
\end{align}
hence its spectral properties are significantly different from the unitary case. Below we briefly outline some of them. 

Liouvillian evolution preserves the trace and hermiticity. Trace-preserving condition can be written as $\bbraket{\mathds{1}|\L|\rho}=0$, where $\mathds{1}$ is the identity matrix and $\rho$ is an arbitrary state. This implies that $\mathds{1}$ is a left eigenvector with zero eigenvalue. In turn, this ensures that there always is at least one right eigenvector $\rho_0$ with zero eigenvalue $\L \rho_0=0$, which is referred to as a steady state. The steady state does not evolve with time and is a counterpart of the ground state for unitary dynamics. 

As a consequence of hermiticity preservation, complex eigenvalues of $\L$ come in conjugated pairs. Indeed, if $\rho_i$ is an eigenstate of $\L$ with eigenvalue $\lambda_i$, i.e. $\L \rho_i=\lambda_i \rho_i$, then by direct conjugation of Eq.~\eqref{Lindblad equation} it follows that $\rho_i^\dagger$ is also an eigenstate with eigenvalue $\bar{\lambda}_i$. 
Real parts of all eigenvalues must be non-positive \cite{Breuer2007, Rivas2011}. In the absence of dissipation, the spectrum of the Liouvillian is purely imaginary.

We restrict our scope to two-dimensional CFTs, where conformal symmetry is described by the Virasoro algebra with generators ${\mathds L}_n$ and commutation relations
\begin{align}
 [\mathds{L}_n,\mathds{L}_m]=(n-m)\mathds{L}_{n+m}+c\frac{n^3-n}{12}\delta_{n+m, 0}   \label{Virasoro} \ .
\end{align}
Here $c$ is the central charge, a key characteristic of a Virasoro algebra. There is a special class of local fields $V_\Delta(z)$, called primary, with straightforward behavior under conformal transformations and simple algebraic correlation functions in the vacuum state, e.g. 
\begin{align}
    \braket{0|V_{\Delta}(z_1)V_{\Delta}(z_2)|0}=\frac{1}{(z_1-z_2)^{2\Delta}} \ . \label{V corr}
\end{align}
where $\Delta$ is the conformal dimension of the field $V$.

At the first glance, generalizing this description to open systems faces technical and conceptual difficulties. Simple form of correlation functions for primary fields relies on the special symmetry properties unique to the vacuum state~$\ket{0}$, namely  $L_{n}\ket{0}=0$ for $n\ge-1$. In open systems the steady state is typically a mixed state, and constructing a mixed state that mimics the symmetry properties of the vacuum state may not look straightforward. More importantly, conformal symmetry is not restricted to spatial dimensions, but manifests itself also in the time correlations. However, defining multi-time correlation functions in open systems can be subtle. Indeed, recall that in the Heisenberg representation operators evolve according to~\cite{Breuer2007} 
\begin{align}
    \partial_t\mathcal{O}=\L^{\dag} \mathcal{O}:=i[H,\mathcal{O}]+\sum_{k} \left(L_k^\dagger \mathcal{O} L_k - \frac12\left\lbrace L_k^\dagger L_k,\mathcal{O}\right\rbrace\right)\ . \label{Lindblad equation op}
\end{align}
Then, since the Liouvillian operator \eqref{Lindblad equation op} is not guaranteed to satisfy the Leibniz rule, i.e. in general $\L^{\dag} \left(\mathcal{O}_1\mathcal{O}_2\right)\neq\left(\L^{\dag} \mathcal{O}_1\right)\mathcal{O}_2+\mathcal{O}_1\left(\L^{\dag}\mathcal{O}_2\right)$, the evolution of composite operators in the Heisenberg picture does not factorize $\left(\mathcal{O}_1\mathcal{O}_2\right)(t)\neq\mathcal{O}_1(t)\mathcal{O}_2(t)$, making multi-time correlators ambiguous.

The Liouvillian evolution operator can be interpreted in terms of doubling of degrees of freedom. Tho copies of the same theory (however with opposite time propagation directions) are coupled by the jump term [the first term in the dissipative part of (\ref{Lindblad equation})] whereas the $\{ \cdot , \cdot \}$ part in Eq.~(\ref{Lindblad equation}) can be combined with the Hamiltonian part thus making it effectively non-Hermitian. Therefore one possible field-theoretical picture could be the following: two non-Hermitian field theories are coupled by the perturbation described by the jump term. In principle, the appearance of non-unitary models in 2D CFT is not an unusual phenomena. The earliest example perhaps is the Lee-Yang CFT~\cite{Fisher1978,Cardy1985} (see also Ref.~\cite{xu2022} for recent developments), namely, the Ising model perturbed by the imaginary magnetic field. This induces the renormalization group flow to the the non-unitary theory with $c=-22/5$. Disordered Dirac fermions are also described by non-unitary CFTs, see e.g. Ref.~\cite{BHASEEN2001465}. Recently, complex CFTs were discussed in the context of weak first order phase transitions \cite{Gorbenko_2018}. Renormalization group flows of non-unitary CFTs were also discussed in Ref.~\cite{Castro-Alvaredo_2017}. Thus, in principle one could study the fixed points of these flows that would correspond to conformal Liouvillians. The above proposal has however one obvious caveat: it is not clear at the moment how to control trace preserving and complete positivity properties along such a flow.  

In this work, we present a description that naturally resolves these tensions and clarifies how exactly conformal symmetry can be realized in Markovian open systems. Our proposal can be summarized as follows.
\begin{itemize}
\item[(i)] Similarly to unitary dynamics, translation-invariant gapless Liouvillian operators give rise to conformal symmetry in open dynamics.

\item[(ii)] Both the Virasoro algebra and the algebra of local fields are represented by superoperators.

\item[(iii)] Spatial conformal symmetry is manifest in correlation functions of superoperators defined as 
\begin{equation}
\Big\langle\!\!\Big\langle{\prod_{i}\pmb{\mathcal{O}}_i}\Big\rangle\!\!\Big\rangle{}_{\rho_0} :=\operatorname{Tr}\left(\prod_i\pmb{\mathcal{O}}_i\right)  {\rho_0}, \label{2pt super}
\end{equation}
where the correlators are evaluated with respect to the steady state $\rho_0$.
    
\item[(iv)] Full space-time conformal symmetry arises in correlation functions of time-dependent superoperators defined as 
\begin{equation}
 \pmb{\mathcal{O}}(t):=e^{-t\L^{\dag}}\pmb{\mathcal{O}}e^{t\L^{\dag}} . \label{super Heisenberg}
\end{equation}
\end{itemize}
The key feature here is to give superoperators the principal role, and reformulate other concepts from the unitary case accordingly\cite{Prosen2008,Kos2017}.  Let us now make a brief connection to the usual operator language.

In some cases, correlation functions of superoperators \eqref{2pt super} can be simply related to correlation functions of ordinary operators. For example, if $\pmb{\mathcal{O}}_i$ act as the left multiplication by ordinary operators, $\pmb{\mathcal{O}}_i\rho=\mathcal{O}_i\rho$, the correlator~\eqref{2pt super} reduces to the standard mixed-state average, e.g.
\begin{align}
    \Big\langle\!\!\Big\langle \pmb{\mathcal{O}}_1\pmb{\mathcal{O}}_2\Big\rangle\!\!\Big\rangle{}_\rho=\braket{\mathcal{O}_1\mathcal{O}_2}_\rho \ .
\end{align}
Hereinafter we denote $\braket{\mathcal{O}}_\rho:=\operatorname{Tr}\mathcal{O}\rho$. In general, however, there will be no such simple reduction.

Similarly, the time evolution \eqref{super Heisenberg} can sometimes be related to the standard Heisenberg evolution \eqref{Lindblad equation op}. In particular, for the two-point function of superoperators $\pmb{\mathcal{O}}_i$ that act as $\pmb{\mathcal{O}}_i\rho=\mathcal{O}_i\rho$, one has the relation
\begin{align}
    \Big\langle\!\!\Big\langle\pmb{\mathcal{O}}_1(t_1)\,\,\pmb{\mathcal{O}}_2(t_2) \Big\rangle\!\!\Big\rangle_{\rho_0}= \braket{\mathcal{O}_1 (t_1-t_2)\mathcal{O}_2}_{\rho_0} \ , \label{2pt from 2pt super}
\end{align}
where $\mathcal{O}(t_1-t_2)$ is a solution to Eq.~\eqref{Lindblad equation op}. Again, no such simple reduction to ordinary correlation function is available in general.

The rest of the paper is essentially a series of examples leading to and illustrating our proposal. In Sec.~\ref{sec basic model} we construct a simple dissipative model with the steady state featuring conformal equal-time correlations. Then, in Sec.~\ref{sec third quantization} we briefly review the formalism of third quantization for quasi-free systems, which will be our key technique allowing for explicit computations of correlations functions and Liouvillian spectrum. In Sec.~\ref{sec basic model revisited} we revisit the model introduced in Sec.~\ref{sec basic model} to rederive its equal-time correlation functions,  as well as define and compute multi-time correlators with the explicit space-time conformal symmetry and associate it with the vanishing Liouvillian gap.  Guided by our basic model, in Sec.~\ref{sec further models} we construct further non-trivial examples of dissipative dynamics with conformal symmetry including (i) a dissipative Luttinger liquid (ii) a model where only the full open dynamics is conformal while the unitary dynamics in the absence of dissipation is not (iii) the fermionic counterpart of our basic model. Finally, in Sec.~\ref{sec outlook} we conclude.

\section{Basic model} \label{sec basic model}

The theory of a free massless boson field $\phi$ in two dimensions, to be defined precisely below, is one of the simplest CFTs \cite{DiFrancesco1997}. Its central charge is $c=1$ and primary fields are the vertex operators $e^{i\alpha\phi}$ with conformal dimensions $\Delta(\alpha)=\alpha^2/4$. Correlators of the vertex operators in the vacuum take the characteristic conformal form~\eqref{vertex corr}. Can a mixed density matrix preserve this conformal form of the correlation functions? If we assume that a density matrix $\rho$ exists such that (i) $\rho$ is Gaussian and (ii) the two-point function of the boson field is the same as the vacuum correlator up to a constant factor $\braket{\phi(x_1)\phi(x_2)}_{\rho}=c \braket{0|\phi(x_1)\phi(x_2)|0}$ then correlation functions of the vertex operators with respect to $\rho$ will only differ from the vacuum correlation functions by an effective rescaling $\phi\to\sqrt{c}\phi$, which preserves the conformal invariance, yet modifies the critical exponents. In this section we will construct a simple dissipative dynamics of the free boson field where the steady state satisfies the required conditions.

\subsection{Unitary free boson} 
The Hamiltonian of the free boson CFT is given by~\cite{DiFrancesco1997, free_boson_H_note}
\begin{align}
H=\sum_{n>0}\omega_n a^\dagger_n a_n ,\label{H free boson}    
\end{align}
where $\omega_n=n\omega$ and $a_n$ are bosonic ladder operators satisfying the canonical commutation relations
\begin{align}
[a_n,a^\dagger_m]=\delta_{nm},\quad [a_n,a_m]=[a_n^\dagger,a_m^\dagger]=0  . \label{aCR}
\end{align}
The chiral boson field defined on a cylinder is constructed out of the ladder operators as follows
\begin{align}
    \phi(x)=i\sum_{n} \frac{1}{\sqrt{2n}} \left(a_n e^{-ik_nx}-a^\dagger_n e^{ik_n x}\right) , \label{phi mode expansion}
\end{align}
where we 
denoted $k_n=kn=\frac{2\pi}{L}n$ and $L$ is the circumference of the cylinder. In the Heisenberg picture, time-dependence of the chiral field is given by
\begin{align}
    \phi(x, t)=e^{iHt}\phi(x)e^{-iHt}=i\sum_{n>0} \frac{1}{\sqrt{2n}} \left(a_n z^{-n}-a^\dagger_n z^n\right) , \label{phi heisenberg}
\end{align}
with the standard notation for the holomorphic coordinate
\begin{align}
z=e^{i(kx+\omega t)} . \label{z def}
\end{align}
Two-point vacuum correlation function of the chiral field reads
\begin{align}
    \braket{0|\phi(z_1)\phi(z_2)|0}=-\frac12\log(z_1-z_2) \ . \label{phi correlator}
\end{align}
As a consequence, normal-ordered vertex operators  are primary fields with the conformal correlation functions
\begin{align}
\bra{0}\prod_i e^{i\alpha_i\phi(x_i, t_i) }\ket{0}=\prod_{i<j}(z_i-z_j)^{\frac12\alpha_i\alpha_j} \ . \label{vertex corr}
\end{align}
We emphasize that this specific form of the correlation functions is closely tied to the conformal symmetry of the theory and to the symmetry properties of the vacuum state with respect to which the correlators are computed.

\subsection{Dissipative free boson}
Let us now introduce the dissipation in such a way that it does not break the conformal symmetry. We examine the time evolution of the density matrix governed by the Lindblad equation \eqref{Lindblad equation} with the free boson Hamiltonian \eqref{H free boson} and the jump operators
\begin{align}
    L_{1n}=\sqrt{\gamma_1\omega} \, a_n,\quad L_{2n}=\sqrt{\gamma_2\omega} \, a_n^\dagger, \, \label{jump def}
\end{align}
where $\gamma_1 = \gamma (\bar{n} + 1)/2, \, \gamma_2 = \gamma \bar{n}/2$. The jump operators correspond to every mode being coupled to its own thermal bath with the average number of excitations~$\bar{n}$ and the decay rate~$\gamma$. The sum in Eq.~\eqref{Lindblad equation} is over all pairs $L_{1n},L_{2n}$ with $n>0$. 

The steady state density matrix $\rho_{0}$ reads
\begin{equation}
    \rho_0=\prod_{n>0}\frac{e^{-\Omega a_n^\dagger a_n}}{1-e^{-\Omega}}, \quad \Omega=\log\frac{\bar{n}+1}{\bar{n}} \ . \label{rho thermal}
\end{equation}
Note that~$\rho_0$ is properly normalized.
If we interpret $ \rho_0 $ in Eq.~\eqref{rho thermal} as a thermal density matrix, then each mode has its own {\it frequency-dependent} temperature 
$T_n = n \omega / \Omega$.
As a result, the density matrix does not introduce any energy scales and is thus compatible with the conformal symmetry. 

Indeed, it is straightforward to check that the only effect of the density matrix~\eqref{rho thermal} on the two-point function of the chiral scalar field is the overall renormalization
\begin{align}   
\braket{\phi(x_1)\phi(x_2)}_{\rho_0}=
(2\bar{n}+1)\braket{0|\phi(x_1)\phi(x_2)|0} \ , \label{thermal to vacuum correlator}
\end{align}
where we denoted \begin{equation}
\braket{\mathcal{O}}_{\rho}=\operatorname{Tr}\rho \mathcal{O}.
\end{equation} 
For a detailed calculation of the correlator~\eqref{thermal to vacuum correlator} see Appendix~\ref{app free boson}. 

Note that since the density matrix \eqref{rho thermal} is Gaussian, the Wick's theorem applies and other correlators in the theory can be derived by a mere rescaling of the fields, e.g.
\begin{align}
\braket{e^{i\alpha\phi(x_1)}e^{-i\alpha\phi(x_2)}}_{\rho_0}=(e^{ik x_1}-e^{ik x_2})^{-\frac12(2\bar{n}+1)\alpha^2}. \label{2pt vertex open}
\end{align}
More generally, for the product of the vertex operators one has
\begin{align}
    \braket{\prod_i e^{i\alpha_i\phi(x_i)}}_{\rho_0}=\prod_{i<j}(e^{ik x_i}-e^{ik x_j})^{\frac12(2\bar{n}+1)\alpha_i\alpha_j}. \label{npt vertex open}
\end{align}
A remarkable conclusion is that the standard conformal form of the vertex correlators is preserved by the steady state \eqref{rho thermal}, although the critical exponents are rescaled. 

Eqs.~\eqref{2pt vertex open} and \eqref{npt vertex open} are equal-time correlators. Replacing $e^{ikx_i}$ by the complex variables $z_i$ from Eq.~\eqref{z def} formally accounts for the unitary time evolution \eqref{phi heisenberg}. Physically this can be realized by preparing the system in the steady state $\rho_0$, then switching off the interactions with the environment. One may call it a dissipative quench. However, we are interested in the full-fledged dissipation theory with the time evolution governed by a Liouvillian with dissipation. We will address this question Sec.~\ref{sec multi-time}, after introducing the formalism of third quantization, which provides a natural framework to clarify the structure of this model and go beyond.

\section{Third quantization of quasi-free systems} \label{sec third quantization}
Quasi-free systems feature a quadratic Hamiltonian and linear jump operators with respect to the ladder operators. In this case, one can represent the Liouvillian as a quadratic form of ladder superoperators, having the standard commutation relations \cite{Prosen2008, Prosen2010}. The formalism of third quantization will be our main technical tool to address conformal symmetry in open quasi-free systems. In this section, we will briefly review the framework following the notation of Ref.~\cite{Barthel2022}.

\subsection{Quadratic form}
For a system of bosonic modes $a_j$ satisfying the canonical commutation relations \eqref{aCR} we introduce their Hermitian linear combinations 
\begin{equation} \label{Majorana_basis}
    w_{j,+}=\frac1{\sqrt2}\left(a_j+a_j^\dagger\right), \qquad
    w_{j,-}=\frac{i}{\sqrt2}\left(a_j - a_j^\dagger\right). 
\end{equation}
The Hamiltonian and the jump operators in the rotated basis read
\begin{align}
    &H = \sum_{i,j,\mu,\nu=\pm} H_{i\mu,j\nu}w_{i\mu}w_{j\nu},\\
    &L_k = \sum_{i,\mu=\pm} L_{k, i\mu}w_{i\mu} \ .
\end{align}
By a slight abuse of notation, we will use the same letters $H$ and $L$ both for the operators and their components. Because the modes~$w_{j, \pm}$ are Hermitian, the Hamiltonian matrix $H$ can always be chosen to be real and symmetric, $H=H^*=H^T$, whereas the vectors $L_k$ are arbitrary.

Now introduce ladder superoperators ($\mu=\pm$)
\begin{align}
\b_{j\mu}\rho=\frac1{\sqrt2}\left\{w_{j\mu},\rho\right\},\quad \b'_{j\mu}\rho=\frac{i \mu}{\sqrt2}[w_{j,-\mu}, \rho] \label{b def} \ ,
\end{align}
which form two independent (mutually commuting) sets of bosonic ladder operators
\begin{align}
 [\b_{i\mu},\b'_{j\nu}]=\delta_{ij}\delta_{\mu\nu},\quad [\b_{i\mu},\b_{j\nu}]=[\b'_{i\mu},\b'_{j\nu}]=0 \label{b comm}\ .
\end{align}
It corresponds to the doubling of the degrees of freedom, when we describe a system with a density matrix.

However, note that the superoperator $\b'_{i \mu}$ is {\it not} the Hermitian conjugate of~$\b_{i \mu}$ . In other words, Eq.~\eqref{b def} is simply a generating set of the superoperator algebra, satisfying ladder commutation relations. The Liouvillian \eqref{Lindblad equation} can be written in terms of ladder superoperators as
\begin{align}
    \L= \sum \b^{\prime}_{i\mu} X_{i\mu,j\nu}\b_{j\nu} + \sum \b'_{i\mu}Y_{i\mu,j\nu}\b'_{j\nu} \ , \label{liouvillian b}
\end{align}
with the matrices $X, Y$ given by
\begin{align}
    X=-2J \Bigl( H-\frac12 \operatorname{Im}\{B\} \Bigr),\qquad Y=-J \operatorname{Re}\{B\} J , \label{XY def}
\end{align}
where $B$ is a Hermitian and positive-semidefinite matrix, explicitly given by
\begin{align}
    B=\sum_{k}L_k L_k^\dagger \ . \label{B def}
\end{align}
In Eq.~\eqref{XY def} we denoted by $J$ the symplectic form for $n$ bosonic modes, which reads
\begin{align}
J=i\sigma_y\otimes \mathds{1}_n=\begin{pmatrix}0&\mathds{1}_n\\-\mathds{1}_n&0\end{pmatrix} \label{J def}.    
\end{align}
We assume that basis modes are ordered so that e.g. $\b=\begin{pmatrix}\b_{1+}\b_{2+}\dots \b_{1-}\b_{2-}\dots\end{pmatrix}$.

Using the superoperator formalism we were able to represent the Liouvillian as a quadratic form with respect to ladder superoperators $\b, \b'$. It means that we can now ignore the Liouvillian being a superoperator, and treat it using standard methods.

\subsection{Liouvillian in the normal form}
The Liouvillian \eqref{liouvillian b} can be brought to the Jordan normal form (JNF) whenever there is a real symmetric matrix $\Gamma=\Gamma^*=\Gamma^T$ that satisfies 
\begin{align}
    X\Gamma+\Gamma X^T+Y=0,  \label{steady state eq}
\end{align}
where $X$ and $Y$ are given by Eq.~(\ref{XY def}) and $\Gamma$ is the covariance matrix of the steady state, $\Gamma_{i\mu,j\nu}=\frac12\braket{w_{i\mu}w_{j\nu}+w_{i\nu}w_{j\mu}}_{\rho_0}$. However, this is not important for our purposes.

Then, the Liouvillian JNF coincides with JNF of the matrix $X$. Indeed, let $\xi$ be the JNF of $X$, and let $S$ represent the corresponding similarity transformation
\begin{align}
 X=S\xi S^{-1} \label{X from xi}  \ . 
\end{align}
Then, in terms of the superoperators
\begin{align}
    \d'=S^T\b',\quad \d=S^{-1}\b-2S^{-1}\Gamma \b' \label{d from b}
\end{align}
the Liouvillian becomes
\begin{align}
    \L=\d' \xi \d \ , \label{liouvillian d}
\end{align}
while the canonical commutation relations \eqref{b comm} are preserved, i.e. $[\d_{i\mu},\d'_{j\nu}]=\delta_{ij}\delta_{\mu\nu}$.

\subsection{Steady state and correlation functions}
The steady state $\rho_0$ of the Liouvillian~\eqref{liouvillian d} is simultaneously annihilated by all operators $\d_{i\mu}$, i.e. $\d_{i\mu}\rho_0=0$. The steady state can also be reconstructed from this condition \cite{Barthel2022}, but this is unnecessary for our purposes. Note that the superoperators $\b$ and $\b'$ are not Hermitian conjugated to each other with respect to the Hilbert-Schmidt inner product \eqref{Hilbert-Schmidt}. Instead one has $\left(\b_{i\mu}\right)^\dagger=\b_{i\mu}$, $\left(\b'_{i\mu}\right)^\dagger=-\b'_{i\mu}$.

Note also that since $\b'$ acts as a commutator \eqref{b comm}, it follows $\b'_{i\mu}\mathds{1}=0$ and hence by linearity \eqref{d from b} also $\d'_{i\mu}\mathds{1}=0$, where $\mathds{1}$ is the identity matrix. This leads us to identify $\rho_0$ with the right vacuum and $\mathds{1}$ with the left vacuum. The existence of different left and right vacuums is natural, because the state space of density matrices is not a Hilbert space. Density matrices form a convex subset of the trace class operator space. Its dual space is the space of bounded operators $B(H)$, which is not isomorphic to the space of density matrices. Hence, it appears only natural that the left and right vacuums do not coincide. Subsequently, we introduce the following notion of correlation functions for superoperators
\begin{align}
\Big\langle\!\!\Big\langle \prod_i\pmb{\mathcal{O}}_i
\Big\rangle\!\!\Big\rangle_{\rho_0}
:=
\Big\langle\!\!\Big\langle 
\mathds{1}\Big| \prod_i\pmb{\mathcal{O}}_i \Big| \rho_0 \Big\rangle\!\!\Big\rangle =\operatorname{Tr}\prod_i\pmb{\mathcal{O}}_i\rho_0 \ . \label{def supercor}
\end{align}
Unusual properties under the Hermitian conjugation render representation theory for correlation functions \eqref{def supercor} different from the unitary case. However, for free systems considered in this paper, the correlators can be computed in a purely algebraic fashion, e.g. $\bbraketvac{\d_{i\mu}\d'_{j\nu}}=\delta_{ij}\delta_{\mu\nu}$.

\subsection{Reverse engineering a Liouvillian}

To construct examples with interesting open dynamics it is useful to answer the following question -- does a Liouvillian with the desired properties exist, i.e. does it correspond to some choice of the Hamiltonian and jump operators? A JNF $\xi$ defines a valid Liouvillian if there exists a similarity transformation $S$ and a matrix $X$ such that 
\begin{align}
    \xi = S^{-1}XS, \quad X=-2J\left(H-\frac12 \im\{B\} \right) \label{xi from H} \ .
\end{align}
Here $H$ is the Hamiltonian in the basis~(\ref{Majorana_basis}), which must be real and symmetric. On the other hand, since $B$ is Hermitian, $\im\{B\}$ is real and skew-symmetric. It is not difficult to show that within these constraints $\im\{B\}$ can be made arbitrary by a suitable choice of jump operators. Furthermore, since $J$ is real and invertible, matrix $X$ can in principle be an arbitrary real matrix. This implies that a JNF of any real matrix can define the Liouvillian spectrum. A caveat here is that one must also require the existence of a covariance matrix $\Gamma=\Gamma^*=\Gamma^T$ solving equation \eqref{steady state eq}. Note that the real part of matrix $B$ does not affect the spectrum, but defines the steady state and the diagonal basis \eqref{d from b}.

Another interesting question to ask is whether some inequivalent unitary dynamics, augmented with appropriate dissipation, can lead to the same JNF of the Liouvillian. Equation \eqref{xi from H} can be solved for the Hamiltonian 
\begin{align}
H=\frac14\left(X^TJ-JX\right) \ . \label{H from X}
\end{align}
Under a symplectic transformation $w\to \beta w$ with $\beta^T J\beta=J$ the Hamiltonian is transformed to
\begin{align}
    H\to\beta^T H\beta=\frac14\left(\left(\beta^{-1}X\beta\right)^TJ-J\beta^{-1}X\beta\right) \ ,
\end{align}
which effectively redefines $S\to \beta^{-1}S$. In particular, a canonical transformation does not affect the JNF because it can be absorbed into a similarity transformation. On the other hand, not every similarity transformation can be compensated by a canonical transformation, as we will demonstrate by an explicit example is Sec.~\ref{sec CFT from dissipation}.

\section{Basic model revisited} \label{sec basic model revisited}
We now translate our basic model to the language of third quantization, which will allow to re-derive equal-time correlation functions from a new perspective, explicitly describe the spectrum of the Liouvillian, and address multi-time correlators.

\subsection{Diagonal form of the Liouvillian}
Following the procedure outlined in Sec.~\ref{sec third quantization}, the Liouvillian for the free boson with the jump operators~\eqref{jump def} can be diagonalized. Details are delegated to Appendix~\ref{app free boson}, the result reads
\begin{equation}
    \L=\sum_{n}\left(-\frac{\gamma}{4}+i\right)\omega_n \d'_{n-}\d_{n-} +\sum_{n}\left(-\frac{\gamma}4-i\right)\omega_n \d'_{n+}\d_{n+} \ , \label{L simple diagonal}
\end{equation}
where
\begin{align}
\begin{split}
    &i\d'_{n-}\rho=[a_n,\rho], \quad 
    i\d_{n-}\rho=- \bar{n} a_n^\dagger \rho+(\bar{n}+1)\rho a_n^\dagger \ , \\ 
    &i\d'_{n+}\rho=[a_n^\dagger, \rho], \quad 
    i\d_{n+}\rho=-(\bar{n}+1)a_n\rho+\bar{n}\rho a_n\ . 
    \label{d simple} 
\end{split}
\end{align}
One can check directly that these operators obey the canonical commutation relations and that $\d_{n\pm}$ annihilate the steady state \eqref{2pt rho0}. Note also that $\left(\d_{n\pm}\rho\right)^\dagger=\d_{n\mp}\rho$ and similarly $\left(\d'_{n\pm}\rho\right)^\dagger=\d'_{n\mp}\rho$ ensuring that the Liouvillian preserves the hermiticity.

Algebraically, the Liouvillian \eqref{L simple diagonal} is a sum of two independent free boson theories with complex spectra conjugated to each other. There are twice as many degrees of freedom in the Liouvillian because the underlying state space is that of density matrices instead of pure states. For $\gamma=0$ the spectrum of the Liouvillian is purely imaginary, which corresponds to the unitary dynamics. The real part of the spectrum is negative in the physical regime $\gamma>0$, as it should be. Note also that the dissipative spectrum is explicitly gapless in the thermodynamic limit, in line with the expected relation between conformal symmetry and the Liouvillian spectrum.

\subsection{Virasoro algebras}
The theory described by the Liouvillian~\eqref{L simple diagonal} features not one but two $c=1$ Virasoro algebras with the generators given by the standard relation
\begin{equation}
\label{dissipative virasoro}
    \mathds{L}_{n\mu}=\sum_{m\in\mathbb{Z}}m:\pmb{d}_{(n-m)\mu}\pmb{d}_{m\mu}: \ ,
\end{equation}
where we denoted $\pmb{d}_{-n}=\pmb{d}'_n$ for $n>0$. The two algebras with different $\mu=\pm$, mutually commute. However, they do not combine into $c=2$ algebra in the same way two unitary free bosons do. The Liouvillian~\eqref{L simple diagonal} can be written as 
\begin{align}
 \L=-\frac{\gamma \omega}{4}\left(\mathds{L}_{0-}+\mathds{L}_{0+}\right)+i\omega\left(\mathds{L}_{0-}-\mathds{L}_{0+}\right)   
\end{align}
and, hence is not simply a sum of $L_{0+}$ and $L_{0-}$. The fact that non-trivial coefficients are allowed is a peculiarity of the open dynamics.

\subsection{Equal-time correlators}
It is natural to introduce the following chiral superfields by analogy with Eq.~\eqref{phi mode expansion}
\begin{align}
    \pmb{\phi}_{\pm}(x)=i\sum_{n>0}\frac1{\sqrt{2n}}\left(e^{-ik_nx}\d_{n\pm}-e^{ik_nx}\d'_{n\pm}\right) \ . \label{superfield}
\end{align}
Equal-time correlators of vertex operators involving these superfields then have the usual conformal properties, e.g.
\begin{align}
    \bbraket{e^{i\alpha \pmb{\phi}_{\mu}(x_1)}e^{-i\alpha \pmb{\phi}_{\nu}(x_2)}}_{\rho_0}=\delta_{\mu\nu} \left(e^{ikx_1}-e^{ikx_2}\right)^{-\frac12\alpha^2} \ .
\end{align}
Naturally, we will refer to the fields like $e^{i\alpha\pmb{\phi}}$ as primary.

We can use this observation to re-derive \eqref{2pt vertex open} in a simple way. Left multiplication by $a_n$ or $a_n^\dagger$ can be expressed via superoperators \eqref{d simple} as follows
\begin{align}
\begin{split}
    &a_n\rho=\left[-i\d_{n+}-i\bar{n}\d'_{n-}\right]\rho \ ,\\
    &a^\dagger_n\rho=\left[i\d_{n-}+i(\bar{n}+1)\d'_{n+}\right]\rho \label{a from d} \ .
\end{split}
\end{align}
These relations can be promoted to the level of the fields
\begin{align}
\begin{split}
    &\overrightarrow{\phi}(x)\rho=\left[-i\overrightarrow{\pmb{\phi}}_+(x)+i\bar{n}\overleftarrow{\pmb{\phi}}_-(-x)\right]\rho \ ,\\
    &\overleftarrow{\phi}(x)\rho=\left[i(\bar{n}+1)\overleftarrow{\pmb{\phi}}_+(x)-i\overrightarrow{\pmb{\phi}}_-(-x)\right]\rho \ .       
\end{split}\label{phi from phi super}
\end{align}
Here $\overleftrightarrow{\phi}, \overleftrightarrow{\pmb{\phi}}_{\pm}$ are creation and annihilation parts of the ordinary full chiral fields and superoperator full chiral fields, respectively, see Appendix~\ref{app aux} for precise definitions. Therefore,
\begin{multline}
    \braket{e^{i\alpha\phi(x_1)}  e^{-i\alpha\phi(x_2)}}_{\rho_0}= \bbraket{e^{\alpha \overrightarrow{\pmb{\phi}}_+(x_1)}e^{\alpha(\bar{n}+1) \overleftarrow{\pmb{\phi}}_+(x_2)}}_{\rho_0} \\
    \times \bbraket{e^{\alpha \overrightarrow{\pmb{\phi}}_-(-x_1)}e^{\alpha\bar{n}\overleftarrow{\pmb{\phi}}_-(-x_2)}}_{\rho_0}\, \label{2pt vertex re-derived}
\end{multline}
 reproduces Eq.~\eqref{2pt vertex open} in view of relations \eqref{2pt rho0}. The correlator factorizes since $\pmb{\phi}_+$ commutes with $\pmb{\phi}_-$. Also, negative signs of coordinates $-x_i$ in the second correlator only produce trivial factors and can be ignored here, but will become important when we generilize to the time-dependent correlators.

\subsection{Multi-time correlators} \label{sec multi-time}
It is natural to ascribe time-dependence to superoperators by $\pmb{\mathcal{O}}(t)=e^{-t\L^{\dag}}\pmb{\mathcal{O}}e^{t\L^{\dag}}$, and define multi-time correlation functions accordingly
\begin{align}
    \bbraket{\pmb{\mathcal{O}}_1(t_1)\dots \pmb{\mathcal{O}}_n(t_n)}_{\rho} \ . \label{time correlator def}
\end{align}
For a conformal Liouvillian with primary superoperators $\pmb{\mathcal{O}}_i$ the resulting correlation function in the steady state will respect the full space-time conformal symmetry. For example, in our current simple model
\begin{align}
\bbraket{e^{i\alpha \pmb{\phi}_{\pm}(x_1, t_1)}e^{-i\alpha \pmb{\phi}_{\pm}(x_2, t_2)}}_{\rho_0} = \left(z^{\pm}_1-z_2^{\pm}\right)^{-\frac{\alpha^2}{2}} \ , \label{2pt pm time dependent}
\end{align}
where we denoted
\begin{align}
    z^\pm=e^{\gamma \omega t/4}e^{ikx\pm i\omega t} \ . \label{zpm def}
\end{align}
These relations are a direct consequence of
\begin{align}
\begin{split}
     \d'_{n\pm}(t)=e^{-t\L^{\dag}}\d'_{n\pm}e^{t\L^{\dag}}=e^{-\left(\frac14\gamma\mp i\right)\omega_n t}\d'_{n\pm} \ ,\\ 
     \d_{n\pm}(t)=e^{-t\L^{\dag}}\d_{n\pm}e^{t\L^{\dag}}=e^{\left(\frac14\gamma\mp i\right)\omega_n t}\d_{n\pm} \ .
\end{split}     
\end{align}

Using connection between the chiral field and primary superfields \eqref{phi from phi super} it is straightforward to compute a time-dependent generalization of \eqref{2pt vertex re-derived}
\begin{widetext}
\begin{equation}
   \bbraket{e^{i\alpha\phi(x_1,t_1)}e^{-i\alpha\phi(x_2,t_2)}}_{\rho_0} 
   = \left(z_1 e^{\frac{\gamma}{4}t_1}-z_2 e^{\frac{\gamma}{4}t_2}\right)^{ -  \alpha^2(\bar{n}+1) / 2 } 
   \left(z_1 e^{-\frac{\gamma}{4}\omega t_1}-z_2 e^{-\frac{\gamma}{4}\omega t_2}\right)^{- \alpha^2\bar{n} / 2 } \ . \label{2pt time}
\end{equation}
\end{widetext}
Here $z_i$ are the standard holomorphic coordinates \eqref{z def} and vertex operators $e^{i\alpha\phi}$ in the original correlator should be understood as superoperators acting by left multiplication. Note that different time-dependence
in $z^{\pm}$ \eqref{zpm def} combines with the opposite $x$-dependence in \eqref{2pt vertex re-derived} to produce holomorphic dependence on $z_i$ in both terms. 

Equation \eqref{2pt time} can also be derived from representation \eqref{2pt from 2pt super}. In the Heisenberg picture, the ladder operators evolve as $a_n(t)=e^{-\frac{\gamma}{4}\omega_nt}a_n, a_n^\dagger(t)=e^{-\frac{\gamma}{4}\omega_nt}a_n^\dagger$ resulting in
\begin{align}
    \phi(x,t)=\rphi\left(z e^{\frac{\gamma}4\omega t}\right)+\lphi\left(z e^{-\frac{\gamma}4\omega t}\right) \ .
\end{align}
It follows that
\begin{equation}
\begin{aligned}
    &\braket{e^{i\alpha \phi(x_1, t_1-t_2)}e^{-i\alpha \phi(x_2)}}_{\rho_0} \\
    &= \braket{e^{i\alpha \rphi(x_1, t_1-t_2)}e^{-i\alpha \lphi(x_2)}}_{\rho_0}\\
    &\times \braket{e^{i\alpha \lphi(x_1, t_1-t_2)}e^{-i\alpha \rphi(x_2)}}_{\rho_0}
\end{aligned}
\end{equation}
is equal to Eq.~\eqref{2pt time} in view of relations~\eqref{2pt rho0}.

\section{Further examples} \label{sec further models}

In this section we consider more examples of dissipative models with conformal Liouvillian dynamics. Each model is a simple quasi-free system and illustrates the general phenomenon in a new situation. 

\subsection{Dissipative Luttinger liquid}
Let us revisit a model introduced in Ref.~\cite{Bacsi2020}, which is defined by the Hamiltonian
\begin{equation}
    H=\sum_{n\in\mathbb{Z}} \omega_n a^\dagger_n a_n,
\end{equation}
and the jump operators
\begin{gather}
L_{n1}=\sqrt{\omega_n}\left(\sqrt{\gamma_1}a_n+\sqrt{\gamma_2}a_{-n}^\dagger\right)\\ L_{n2}=\sqrt{\omega_n}\left(\sqrt{\gamma_1}a_{-n}+\sqrt{\gamma_2}a_{n}^\dagger\right) , \label{setup luttinger}
\end{gather}
where $\gamma_1 = \gamma (\bar n +1)/2, \, \gamma_2 = \gamma \bar n/2$.

It is similar in spirit to our basic example considered in the previous section, although the Hamiltonian includes both left- and right-moving modes and the jump operators mix left and right modes of equal energy. Following the procedure of third quantization, this model can be diagonalized quite similarly to our basic example (details are presented in Appendix~\ref{app luttinger})
\begin{multline}
\L=\!\sum_{n\in\mathbb{Z}}\Bigl(-\frac12\gamma_{12}+i \Bigr)\omega_n \d_{n-}'\d_{n-} \\ 
+ \sum_{n\in\mathbb{Z}} \Bigl( 
-\frac14\gamma-i
\Bigr) \omega_n \d_{n+}'\d_{n+}, \label{L luttinger diagonal}
\end{multline}
where the superoperators act on the density matrix~$\rho$ as
\begin{align}
\begin{split}
&i\d'_{n-}\rho=[a_n,\rho]\ ,\\
&i\d_{n-}\rho=-\bar n a_n^\dagger\rho+(1 + \bar n)\rho a_n^\dagger-\frac{\sqrt{\bar n (\bar n + 1)}[a_{-n},\rho]}{\bar n - 4i/\gamma}\ ,\\
&i\d'_{n+}\rho=[a^\dagger_n,\rho]\ ,\\
&i\d_{n+}\rho=-(\bar n + 1) a_n\rho+\bar n\rho a_n-\frac{\sqrt{\bar n (\bar n +1)}[a^\dagger_{-n},\rho]}{\bar n + 4i/\gamma}\ . \label{d luttinger}
\end{split}
\end{align}
A key observation made in Ref.~\cite{Bacsi2020} is that equal-time correlation functions of the right-moving fermion operators $\Psi_R(x)=e^{i\sqrt{2}\phi(x)}$ preserve their conformal form in the steady state, albeit with modified critical exponents. It takes a simple generalization of our basic example to confirm this fact. Left multiplication by creation and annihilation modes can be expressed as
\begin{align}
&a_n\rho =-i\left[\d_{n+}+ \bar n \d'_{n-}+\frac{\sqrt{\bar n (\bar n + 1)}}{\bar n +4i/\gamma}\d'_{(-n)+}\right]\rho \ ,\\
&a_n^\dagger\rho=i\left[\d_{n-}+(\bar n + 1)\d'_{n+}+\frac{\sqrt{\bar n (\bar n + 1)}}{\bar n - 4i/\gamma}\d'_{(-n)-}\right]\rho \ .
\end{align}
The first two terms in each of these expressions coincide precisely with Eq.~\eqref{a from d}. Note that the last terms in both $a_n\rho$ and $a_n^\dagger\rho$ contain only creation superoperators with negative index $\d'_{(-n)\pm}$, which commute with each other and all other operators in both $a_n$ and $a_n^\dagger$. Hence, these terms do not contribute to the steady-state correlators of the chiral vertex operators. Therefore, decomposition \eqref{phi from phi super} still holds in this model up to the terms that do not affect correlation functions of $\Psi_R$ alone. The result \eqref{2pt vertex open} then applies for the correlation function $\braket{\Psi_R(x_1)\Psi_R(x_2)}_{\rho_0}$ and matches precisely the correlator found in Eq.~\cite{Bacsi2020} upon identification $\gamma_1=\eta^2, \gamma_2=1$.

\subsection{Conformal symmetry from dissipation} \label{sec CFT from dissipation}
In the examples we considered so far the unitary dynamics was conformal, and the additionally introduced dissipation preserved the symmetry. It is interesting to ask if an open evolution can be conformal when the original unitary dynamics is not. We answer this question affirmatively, by engineering a simple model. Our construction is based on the observation that the Liouvillian spectrum does not fully fix the Hamiltonian dynamics. Indeed, different choices of the similarity transformation $S$ in Eq.~\eqref{X from xi} correspond to the same Liouvillian spectrum $\xi$, but to different unitary dynamics determined via Eq.~\eqref{H from X}.

We construct a model involving two interacting modes~$a_{\pm n}$. Apparently, for a single mode a similar example does not exist, i.e. the similarity transformation can always be compensated by a symplectic basis change. Consider the following Hamiltonian
\begin{multline}
 H = \sum_{n>0}\left[\omega_n\left(a_n^\dagger a_n+a_{-n}^\dagger a_{-n}\right)\right.\\
 \left. +\frac{\mu(n)\omega_n}{2}\left(a^\dagger_na^\dagger_{-n}+a_{n}a_{-n}\right)\right]  \label{H mu}
\end{multline}
and the jump operators 
\begin{align}
\begin{split}
&L_{1n}=\frac{\sqrt{\omega_n}}{2\sqrt{2}}\left[\mu(n)\left(a_n+a_n^\dagger\right)-2i\left(a_{-n}+a_{-n}^\dagger\right)\right] \ ,\\
&L_{2n}=\frac{\sqrt{\omega_n}}{2\sqrt{2}}\left[\mu(n)\left(a_n-a_n^\dagger\right)-2i\left(a_{-n}-a_{-n}^\dagger\right)\right] \ ,\\
&L_{3n}=\sqrt{\omega_n}\left(\sqrt{\gamma_1}a_n+\sqrt{\gamma_2}a_n^\dagger\right) \ ,\\
&L_{4n}=\sqrt{\omega_n}\left(\sqrt{\gamma_1}a_{-n}+\sqrt{\gamma_2}a_{-n}^\dagger\right) \ .
\end{split}
\end{align}
Here $\mu(n)$ is an arbitrary real function. This model is designed for the diagonal form of the Liouvillian not to depend on $\mu(n)$. Indeed, as detailed in Appendix~\ref{app from dissipation} the diagonal Liouvillian is exactly the same as written in Eq.~\eqref{L luttinger diagonal}. Therefore, the Liouvillian dynamics is conformal with the Virasoro algebra realized by~\eqref{dissipative virasoro}. However, in the original unitary model \eqref{H mu} the spectrum obviously depends on $\mu(n)$. This can be seen by computing determinant 
\begin{align}
\operatorname{det}(H_n-\lambda J)=\left(\lambda+\frac{\omega_n^2}{16}\left(4-\mu(n)^2\right)\right)
\end{align}
where $H_n$ is the Hamiltonian for modes $a_n, a_{-n}$ in the basis~(\ref{Majorana_basis}) and $J$ is the symplectic form~\eqref{J def}. This determinant is invariant under symplectic transformations and hence explicit dependence on $\mu(n)$ shows that unitary models are not equivalent. Choosing $\mu(n)$ to depend non-trivially on the energy one can break conformal symmetry of the unitary model without affecting properties of the Liouvillian dynamics.

\subsection{Fermionic example} 

The models considered thus far involve bosonic fields, but constructing examples of open fermionic systems with conformal symmetries in the same way does not pose any significant difficulties. As an illustration we will now present the fermionic counterpart of our basic model. The Hamiltonian and the jump operators are given by
\begin{gather}
    H=\sum_{n\ge\frac12}\omega_n c_n^\dagger c_n, \\
    L_{n1} = \sqrt{\gamma_1\omega_n}c_n,\qquad L_{n2} = \sqrt{\gamma_2\omega_n}c_n^\dagger \ , \label{fermionic model}
\end{gather}
with $\omega_n=\omega n$ and canonical fermionic commutation relations $\{c_n,c_n^\dagger\}=1, c^2=(c^\dagger)^2=0$. The steady state is given by
\begin{equation}
\rho_0=\prod_{n\ge\frac12} \frac{e^{-\Omega c^\dagger_nc_n}}{1+e^{-\Omega}}, \quad \Omega=\log{\frac{\gamma_1}{\gamma_2}}    \ .
\end{equation}
Similarly to the bosonic case, the equal-time correlation functions of chiral fermion operators 
\begin{align}
\psi(x)=ie^{-ikx/2}\sum_{n\ge\frac12}\left(c_{n} e^{-ik_{n} x}-c_{n}^\dagger e^{ik_{n} x}\right) \label{psi def}
\end{align}
are renormalized in the steady state with respect to the vacuum correlators
\begin{multline}
\braket{\psi(x_1)\psi(x_2)}_{\rho_0} = \frac{\gamma_{1}-\gamma_2}{\gamma_1+\gamma_2}\braket{0|\psi(x_1)\psi(x_2)|0} \\
 =\frac{\gamma_{1}-\gamma_2}{\gamma_1+\gamma_2}\frac{1}{e^{i k x_1}-e^{ik x_2}}, \label{2pt fermions}
\end{multline}
due to the identities 
\begin{multline}
\braket{c_n c_n^\dagger}_{\rho_0}=1-\braket{c_n^\dagger c_n}_{\rho_0} = 1/(1 + e^{-\Omega}). 
\end{multline}
We note that in contrast to the bosonic case, renormalization of this correlator does not alter the critical exponents.

Third quantization for fermions is very similar to bosons, yet with important technical distinctions as discussed in Appendix~\ref{app fermion}. In particular, the Liouvillian acts differently on density matrices with even or odd number of fermions. In the even sector for our model one has
\begin{multline}
    \L=\sum_{n\ge\frac12}\left(-\frac{\gamma_1+\gamma_2}{2}+i\right)\omega_n \d'_{n-}\d_{n-} \\
    + \sum_{n\ge\frac12}\left(-\frac{\gamma_1+\gamma_2}{2}-i\right)\omega_n \d'_{n+}\d_{n+} \ , \label{Liouvillian fermion}
\end{multline}
where the jump operators act on the density matrix as
\begin{align}
\begin{split}
i\d'_{n-}\rho=-c_n\rho- \rho_\Pi c_n, \quad i\d_{n-}\rho=\frac{\gamma_2 c_n^\dagger\rho-\gamma_1 \rho_\Pi c_{n}^\dagger}{\gamma_1+\gamma_2} \ ,\\
i\d'_{n+}\rho=c_n^\dagger\rho+\rho_\Pi c^\dagger,\quad i\d_{n+}\rho=\frac{-\gamma_1 c_{n}\rho+\gamma_2 \rho_\Pi c_n}{\gamma_1+\gamma_2}\ . \label{d fermion def}
\end{split}
\end{align}
Here $\rho_\Pi=\Pi \rho \Pi$, with $\Pi=\prod_{n\ge\frac12}(-1)^{c_n^\dagger c_n}$ being the fermion parity operator. In the odd sector the Liouvillian has the same form with $\d'$ and $\d$ interchanged, details are given in Appendix~\ref{app fermion}. 

Similarly to Eq.~\eqref{phi from phi super} we find
\begin{align}
\begin{split}
    &\rpsi(x)\rho=\left[-i\rbpsi_+(x)+i\frac{\gamma_2}{\gamma_1+\gamma_2}e^{-ikx}\lbpsi_-(-x)\right]\rho \ ,\\
    &\lpsi(x)\rho=\left[i\frac{\gamma_1}{\gamma_1+\gamma_2}\lbpsi_+(x)-ie^{-ikx}\rbpsi_-(-x)\right]\rho \ .\label{psi from psi super}
\end{split}
\end{align}
Using this identification the two-point correlator \eqref{2pt fermions} is straightforward to reproduce. Explicit form of the Liouvillian shows that the open dynamics consists of two decoupled free fermion models and thus exhibits the full space-time conformal symmetry.

It would be also interesting to construct a nonunitary fermionic CFT with the central charge $c=-2$. We discuss this research direction in Appendix~\ref{app symplectic}.

\section{Summary and outlook} \label{sec outlook}
In this work, we proposed a general framework for describing conformal symmetry in Markovian open systems and illustrated it with several examples. For quasi-free systems, i.e. systems with quadratic Hamiltonians and linear jump operators, the formalism of third quantization allows us to cast the Liouvillian as a quadratic form of ladder superoperators with the usual canonical commutation relations. Spectral properties of the Liouvillian and correlation functions of superoperators are manifest in this description. Gapless Liouvillians can often be related to simple unitary conformal Hamiltonians, and this correspondence allows for direct construction of many familiar conformal structures for the open systems.

Our analysis was restricted to algebraic properties of conformal Liouvillians which closely mimic those of conformal Hamiltonians. However, the underlying state space and representation theory will apparently be rather different. For one, in contrast to the pure states forming a vector space, the density matrices are only a convex subset therein. Furthermore, creation and annihilation superoperators $\b, \b'$ \eqref{b def} as well as left and right vacuum states are not Hermitian conjugates of each other, which should make unitary representations different. A close connection between dissipation and non-unitarity is expected, but remains to be fully clarified.

All the models we introduced were based on quasi-free systems and led to quadratic Liouvillians. However, the core proposal equally applies to interacting theories, and it would be particularly interesting to construct explicit examples of this kind. A direct approach could start with some CFT and try designing the jump operators, perhaps building them from local primary fields, that would lead to a conformal dissipative theory. Investigating higher-dimensional models appear to be another intriguing possibility. Quite generally, the framework we introduced should allow to generalize the bootstrap approach to CFT \cite{Ferrara1973,Polyakov1974} to open systems, although coming up with explicit solutions might require substantially new methods. 

Taking into the account that in the vicinity of a critical point one can significantly increase measurement precision \cite{Zanardi2008} our findings could potentially be of interest in the field of quantum technologies. 

\acknowledgements

The work of VG is part of the DeltaITP consortium, a program of the Netherlands Organization for Scientific Research (NWO) funded by the Dutch Ministry of Education, Culture and Science (OCW).
The work of AF is supported by the RSF Grant 19-71-10092. 
This work is also supported by is supported by the Priority 2030 program at the National University of Science and Technology “MISIS” under the project K1-2022-027. 

\appendix

\section{Basic bosonic model} \label{app free boson}

\subsection{Chiral field and vacuum correlators}

We write expansion of the chiral boson field as
\begin{align}
\phi(x)=\rphi(x)+\lphi(x) \ ,
\end{align}
where
\begin{align}
    &\overrightarrow{\phi}(x)=i\sum_{n} \frac{1}{\sqrt{2n}} a_n e^{-ik_nx},\\
    &\overleftarrow{\phi}(x)=-i\sum_{n} \frac{1}{\sqrt{2n}} a_n^\dagger e^{ik_nx} \label{phi left right}
\end{align}
are creation and annihilation parts of the chiral field, so that $\overrightarrow{\phi}(x)\ket{0}=0$ and $\bra{0} \overleftarrow{\phi}(x)=0$. The basic vacuum propagator \eqref{phi correlator} can be obtained by a straightforward computation
\begin{equation}
\begin{aligned}
&\braket{ 0 |\phi(x_1)\phi(x_2) | 0} = \braket{ 0 | \rphi(x_1)\lphi(x_2) | 0 }\\
& =\frac12\log\left(e^{ikx_1}-e^{ikx_2}\right) \ , \label{2pt simeq}
\end{aligned}
\end{equation}
Note that we have omitted the rigorous treatment of the zero mode~\cite{DiFrancesco1997} in the equation above, since it is completely standard and not important for our purposes.

\subsection{Correlators in conformal steady state}
In the ensemble~\eqref{rho thermal} correlation functions of the  creation and annihilation parts of the chiral field are modified as follows
\begin{align}
\begin{split}
\braket{\rphi(x_1)\lphi(x_2)}_{\rho_0}=(\bar{n}+1)\braket{0|\rphi(x_1)\lphi(x_2)|0} \ ,\\
\braket{\lphi(x_1)\rphi(x_2)}_{\rho_0}=\bar{n}\langle 0 \big| \rphi(x_2)\lphi(x_1) \big| 0 \rangle \ , \label{2pt rho0}
\end{split}
\end{align}
due to the identities for thermal averages $\operatorname{Tr}a_n^\dagger a_n \rho_0=\operatorname{Tr}a_n a_n^\dagger\rho_0 - 1= 1/ ( e^\Omega - 1 )$. We note that the fields ordering in the second correlator is reversed. The full correlator then reads
\begin{multline}
    \braket{\phi(x_1)\phi(x_2)}_{\rho_0} = (\bar{n}+1)\braket{0|\phi(x_1)\phi(x_2)|0} \\ +\bar{n}\braket{0|\phi(x_2)\phi(x_1)|0}  \\
    = (2\bar{n}+1)\braket{0|\phi(x_1)\phi(x_2)|0}
\end{multline}

The last relation, acquiring~\eqref{thermal to vacuum correlator}, holds due to Eq.~\eqref{2pt simeq} and the fact that interchanging $x_1\leftrightarrow x_2$ is a symmetry of the vacuum correlator (up to an irrelevant additive constant). 

\subsection{Third quantization}
As neither the Hamiltonian nor the jump operators in model \eqref{Lindblad equation} involve interactions between different bosonic modes, the problem splits into a direct sum of single-mode problems indexed by $n$. The matrix $H_n$ and the vectors $L_{1n}, L_{2n}$ defining the Hamiltonian and the jump operators in the basis~(\ref{Majorana_basis}) for the $n$th mode are
\begin{gather}
  H_n = \frac{\omega}{2}\mathds{1},\\ 
  L_{1n} =\sqrt{\frac{\gamma_1\omega_n}{2}}\begin{pmatrix}1\\-i\end{pmatrix},\quad L_{2n}=\sqrt{\frac{\gamma_2\omega_n}{2}}\begin{pmatrix}1\\i\end{pmatrix}
\end{gather}
Matrices $X_n, Y_n$ defining the quadratic Liouvillian superoperator \eqref{XY def} are
\begin{align}
    X_n=-\frac{\omega_n\gamma}{4}\mathds{1}-i\omega_n \sigma_y, \quad Y_n=\frac{\gamma(2\bar{n}+1)\omega_n}{4}\mathds{1} \ .
\end{align}
Here and below $\mathds{1}$ is the $2\times2$ identity matrix and $\sigma_{x,y,z}$ are the Pauli matrices.

The matrix $X_n$ can be diagonalized as $X=S\xi_n S^{-1}$ with
\begin{align}
    \xi_n = \omega_n\begin{pmatrix}-\frac{\gamma}4-i&0\\0&-\frac{\gamma}4+i\end{pmatrix},\quad S=\begin{pmatrix} -i&i\\1&1\end{pmatrix},
\end{align}
which gives us the Liouvillian spectrum \eqref{L simple diagonal}. The explicit form of diagonal superoperators~\eqref{d simple} can be found from the relation~\eqref{d from b}, by taking into account that the covariance matrix of the steady state~\eqref{steady state eq} is 
\begin{align}
    \Gamma_n=\frac{2\bar{n}+1}{2}\mathds{1} \ .
\end{align}

\subsection{Chiral superfield} \label{app aux}
Similarly to the decomposition \eqref{phi left right} let us split the superfields \eqref{superfield} as $\pmb{\phi}_{\pm}(x)=\overrightarrow{\pmb{\phi}}_{\pm}(x)+\overleftarrow{\pmb{\phi}}_{\pm}(x)$, where
\begin{align}
  &\overrightarrow{\pmb{\phi}}_{\pm}(x)=i\sum_{n} \frac{1}{\sqrt{2n}} \d_{n\pm} e^{-ik_nx},\\
  &\overleftarrow{\pmb{\phi}}_{\pm}(x)=-i\sum_{n} \frac{1}{\sqrt{2n}} \d'_{n\pm} e^{-ik_nx}  \ . 
\end{align}
The key algebraic properties of these fields related to correlator computations are $\overrightarrow{\pmb{\phi}}_{\pm}\rho_0=0$, $\overleftarrow{\pmb{\phi}}_{\pm}\mathds{1}=0$.

\section{Dissipative Luttinger liquid} \label{app luttinger}
Since the jump operators \eqref{setup luttinger} only mix two modes, the problem splits into a direct sum of two-mode problems indexed by $n>0$. For Hamiltonian and jump operators we find the following matrix representations in the basis~(\ref{Majorana_basis})
\begin{align}
  H_n&=\frac{\omega_n}{2}\mathds{1},\\
  L_{1n}&=\sqrt{\frac{\omega_n}{2}}\begin{pmatrix}\sqrt{\gamma_1}\\\sqrt{\gamma_2}\\-i\sqrt{\gamma_1}\\i\sqrt{\gamma_2}\end{pmatrix},\quad 
  L_{2n}=\sqrt{\frac{\omega_n}{2}}\begin{pmatrix}\sqrt{\gamma_2}\\\sqrt{\gamma_1}\\i\sqrt{\gamma_2}\\-i\sqrt{\gamma_1}\end{pmatrix} \ .
\end{align}
The matrices $X_n, Y_n$ defining the quadratic form of the Liouvillian are given by
\begin{align}
\begin{split}
X_n &=-\frac{\gamma_1 - \gamma_2}2\omega_n\,\,\mathds{1}\otimes\mathds{1}-i\omega_n\,\, \sigma_y\otimes\mathds{1}\ , \\
Y_n & =\frac{\gamma_1+\gamma_2}{2}\omega_n\,\,\mathds{1}\otimes\mathds{1}-\sqrt{\gamma_1\gamma_2}\omega_n\,\,\sigma_z\otimes\sigma_x \ .
\end{split}
\end{align}
The matrices $X_n$ are diagonalized by a similarity transformation $X_n=S\xi_nS^{-1}$ with
\begin{align}
\begin{split}
  \xi_n&=-\frac{\gamma_1 - \gamma_2}{2}\omega_n\,\,\mathds{1}\otimes\mathds{1}-i\omega_n\,\, \sigma_z\otimes\mathds{1}, \\ 
  S&=\begin{pmatrix} 0 & -i & 0 & i \\ -i & 0 & i & 0 \\ 0 & 1 & 0 & 1 \\ 1 & 0 & 1 & 0 \end{pmatrix}. \label{luttinger xi}
\end{split}
\end{align}
The spectrum of $\xi$ defines the spectrum of the Liouvillian \eqref{L luttinger diagonal}. To find the explicit form of the diagonal operators \eqref{d luttinger} one needs to use relation \eqref{d from b} and the covariance matrix $\Gamma$ of the steady state \eqref{steady state eq}. The latter can be found to be
\begin{multline}
  \Gamma_n=\frac{\gamma_1+\gamma_2}{2\gamma_{12}}\mathds{1}\otimes\mathds{1}-\frac{\sqrt{\gamma_1\gamma_2}\gamma_{12}}{\gamma_{12}^2+4}\sigma_z\otimes\sigma_x\\
  -\frac{2\sqrt{\gamma_1\gamma_2}}{\gamma_{12}^2+4}\sigma_x\otimes\sigma_x \ .
\end{multline}
Note that the covariance matrices $\Gamma_n$ and the similarity transformation $S$ do not depend on $n$.

\section{CFT from dissipation} \label{app from dissipation}
For each pair of modes $a_n, a_{-n}$ in the model \eqref{H mu} the Hamiltonian in the basis~(\ref{Majorana_basis}) is given by
\begin{align}
H_n=\frac{\omega_n}{2}\mathds{1}+\frac{\mu(n)\omega_n}{4}\sigma_z\otimes \sigma_x\ ,
\end{align}
and the jump operators read
\begin{align}
\begin{split}
L_{1n}=\sqrt{\omega_n}\begin{pmatrix}\frac{\mu(n)}{2}\\-i\\0\\0\end{pmatrix}, \; & L_{3n}=\sqrt{\frac{\omega_n}{2}}\begin{pmatrix}\sqrt{\gamma_1}+\sqrt{\gamma_2}\\0\\-i\sqrt{\gamma_1}+i\sqrt{\gamma_2}\\0\end{pmatrix}, \\
L_{2n}=\sqrt{\omega_n}\begin{pmatrix}0\\0\\-\frac{i\mu(n)}{2}\\1\end{pmatrix}, \; &L_{4n}=\sqrt{\frac{\omega_n}{2}}\begin{pmatrix}0\\\sqrt{\gamma_1}+\sqrt{\gamma_2}\\0\\-i\sqrt{\gamma_1}+i\sqrt{\gamma_2}\end{pmatrix}.
\end{split}
\end{align}
The matrices $X_n,Y_n$ defining the quadratic Liouvillian are 
\begin{align}
\begin{split}
  X_n &= -\frac{\gamma_{12}}{2}\omega_n \mathds{1}\otimes\mathds{1}-i\omega_n \sigma_y \otimes \mathds{1} \\
  &+\frac{\mu(n)\omega_n}{2}\sigma_x\otimes (\sigma_x-i\sigma_y), \\
  Y_n &= \frac{\gamma_1+\gamma_2}{2}\omega_n\mathds{1}\otimes\mathds{1}-\frac{\sqrt{\gamma_1\gamma_2}}{2}\sigma_z\otimes \mathds{1} \\
  &+\frac{\mu^2(n)}{4}\omega_n\mathds{1}\otimes(\mathds{1}+\sigma_z)+\frac12\omega_n \mathds{1}\otimes (\mathds{1}-\sigma_z) \ .
\end{split}
\end{align}
The spectrum of the Liouvillian $\xi_n$ coincides with that in Eq.~\eqref{luttinger xi} and the similarity transformation $S_n$ reads
\begin{align}
S_n=\left(
\begin{array}{cccc}
 0 & -i & 0 & i \\
 -i & i \mu(n)  & i & -i \mu(n)  \\
 0 & 1 & 0 & 1 \\
 1 & 0 & 1 & 0 \\
\end{array} 
\right) \ .
\end{align}

\section{Basic fermionic model} \label{app fermion}
\subsection{Third quantization}
Third quantization for fermions is analogous to bosons albeit with some technical differences. For $n$ fermionic modes $c_1, c_1^\dagger,\dots, c_n,c_n^\dagger$ we define $2n$ Majorana modes by
\begin{equation}
    w_{k}= \frac{c_k+c_k^\dagger}{\sqrt{2}}, \quad w_{k+n}=i\frac{c_k-c_k^\dagger}{\sqrt{2}}, \qquad 1\le k\le n
\end{equation}
so that $\{w_i,w_j\}=\delta_{ij}$. The state space can be split into the even and odd sectors according to the eigenvalue of the fermion parity operator $\Pi=\prod_k (-1)^{c_k^\dagger c_k}$. Then, the quadratic Liouvillian in the even sector is given by
\begin{equation}
\L=\b'X\b+i\b'Y\b' \label{L fermion} \ ,
\end{equation}
with ladder superoperators defined as
\begin{equation}
\b'_{i}=\frac{1}{\sqrt{2}}\left(w_i\rho+\rho_\Pi w_i\right),\quad \b_{i}=\frac{1}{\sqrt{2}}\left(w_i\rho-\rho_\Pi w_i\right),
\end{equation}
so that they obey the canonical commutation relations $\{\b'_i,\b_j\}=\delta_{ij}$, $\{\b'_i,\b'_j\}=\{\b_i,\b_j\}=0$. Here $\rho_\Pi=\Pi \rho \Pi$.

In the odd sector the Liouvillian has the same form with $\b'$ and $\b$ interchanged. The matrices $X$ and $Y$ are defined by
\begin{align}
    X=-2iH-\operatorname{Re}\{ B \},\quad Y = \operatorname{Im}\{B\}
\end{align}
with $H=H^\dagger=-H^T$ being the fermionic Hamiltonian in the Majorana basis and the matrix $B$ defined in Eq.~\eqref{B def}. The steady-state covariance matrix $\Gamma$ is defined as a solution to Eq.~\eqref{steady state eq} satisfying $\Gamma=\Gamma^*=-\Gamma^T$. Note that for the fermion systems the solution always exists.

The Liouvillian \eqref{L fermion} can be brought to the JNF $\xi$ by a similarity transformation $X=S\xi S^{-1}$ and it reads
\begin{align}
\L = \d' \xi \d.
\end{align}
The similarity transformation is induced by the linear transformation of ladder superoperators preserving the canonical commutation relations
\begin{align}
\d = S^{-1}\b-2iS^{-1}\Gamma \b,\quad \d'=S^T \b' \ . \label{d from b ferm}
\end{align}

\subsection{Explicit computations}
For the model \eqref{fermionic model} we find
\begin{align}
  H_n&=\frac{\omega_n}{2}\sigma_y,\\
  L_{1n}&=\sqrt{\frac{\omega_n\gamma_1}{2}} \begin{pmatrix}1\\ -i\end{pmatrix},\quad L_{2n}=\sqrt{\frac{\omega_n\gamma_2}{2}} \begin{pmatrix}1\\i\end{pmatrix}\ .
\end{align}
and hence
\begin{align}
    X_n &= -\frac{\gamma_1+\gamma_2}{2}\omega_n\mathds{1}-i\omega_n\sigma_y,\quad Y_n=i\frac{\gamma_{1}-\gamma_2}{2}\omega_n\sigma_y,\\ 
    \Gamma_n &=i\frac{\gamma_1-\gamma_2}{2(\gamma_1+\gamma_2)}\sigma_y  \ . 
\end{align}
Diagonal form of $\xi_n$ of $X_n=S\xi_n S^{-1}$ and the corresponding similarity transformation are given by
\begin{align}
    \xi_n=-\frac{\omega_n(\gamma_1+\gamma_2)}{2}\mathds{1}-i\omega_n \sigma_y,\quad S=\begin{pmatrix}-i&i\\1&1\end{pmatrix} \ .
\end{align}
Explicit form of diagonal superoperators \eqref{d fermion def} follows from relation \eqref{d from b ferm}.

\subsection{Fermionic fields}
We introduce creation and annihilation parts of the standard fermionic field \eqref{psi def}
\begin{align}
\begin{split}
    &\rpsi(x)=ie^{-ikx/2}\sum_{n\ge\frac12}c_{n}e^{-ik_{n}x},\\
    &\lpsi(x)=-ie^{-ikx/2}\sum_{n\ge\frac12}c_{n}^\dagger e^{ik_{n}x}
\end{split}
\end{align}
so that we have $\psi(x)=\rpsi(x)+\lpsi(x)$ and $\rpsi(x)\ket{0}=\bra{0}\lpsi(x)=0$. 

Similarly, we introduce their superfield counterparts
\begin{align}
\begin{split}
    &\rbpsi_{\pm}(x)=ie^{-ikx/2}\sum_{n\ge\frac12}\d_{n\pm}e^{-ik_nx},\\
    &\lbpsi_{\pm}(x)=-ie^{-ikx/2}\sum_{n\ge\frac12}\d'_{n\pm} e^{ik_{n}x} \ ,
\end{split}
\end{align}
which feature in relations \eqref{psi from psi super} and satisfy $\rbpsi_{\pm}(x)\rho_0=0$ and $\mathds{1}\lbpsi_{\pm}(x)=0$.

\section{Symplectic fermions in Lindblad framework} \label{app symplectic}

We aim to investigate a logarithmic CFT in the context of Lindbladian dynamics. Naively, it appears that a logarithmic CFT being non-unitary should describe a dissipative system. On the other hand, a general dissipative system exhibits a completely positive time evolution generated by a Lindbladian. However, an abstract Hilbert space representing the state space of a conformal field theories lacks essential structure. In Lindbladian dynamics we work with a state space realized by density matrices, allowing the definition of positive density matrices -- a convex subset of all density matrices. Starting with an abstract Hilbert space, we can select a certain class of linear functionals to define positivity, but multiple ways of doing it exist. Moreover, we need a specific linear functional to realize trace operation on density matrices. We can also apply Hermitian conjugation to the operators, which is not inherently defined on a Hilbert space. Without defining density matrix positivity and Hermitian conjugation by hand it does not make sense to talk about whether the evolution is completely positive, preserves the Hermitian density matrices, or preserves the trace.

Subsequently, the Hilbert space of typical CFTs lacks this structure. Therefore, we need to add it by hand. The general question of possible density matrix structures consistent with the Hilbert space is too complex to discuss in this appendix. Thus, we focus on a specific case, searching for representations $\pi$ of a CFT algebra in a density matrix space. Naturally, different representations will emerge. However, it is not evident that the chosen density matrix structure will be sufficiently ``good''. In other words, it is not guaranteed that the time evolution dynamics governed by $ \pi(\mathds{L}_0) $ will be completely positive and/or preserve the hermitian density matrices.

In this appendix we consider this problem illustrated by symplectic fermion model \cite{Creutzig2013}. Being one of the simplest logarithmic CFT, it is a good toy model to illustrate the issues we are talking about. The algebra of symplectic fermions can be defined by two currents 
\begin{align}
    J^{\pm}(z)=\sum_{n}\frac{J^{\pm}_n}{z^n}
\end{align}
with the commutation relations 
\begin{align}
    \{J^{+}_n, J^{-}_m\}=n\delta_{n+m,0}, \qquad \{J^{+}_n, J^{+}_m\}=\{J^{-}_n, J^{-}_m\}=0 . \label{sf algebra}
\end{align}

The zero Virasoro generator has the following form
\begin{align}
\mathds{L}_0=J^-_0J^+_0+\sum_{n\ge1}\left(J^-_{-n}J^+_n-J^+_{-n}J^-_{n}\right).   \label{L0 integer}
\end{align}
where $n$ runs over integers in the case of the periodic boundary conditions.

As symplectic fermions realize a logarithmic CFT due to the zero mode being non-diagonalizable, below we examine only the zero mode. It allows us to avoid dealing with infinite-dimensional space and make the argument more simple and transparent. 

We employ the method of third quantization \cite{Prosen2008} to realize symplectic fermions as superoperators acting on density matrices of ordinary Majorana fermions $\psi, \bar\psi$. Hereinafter, we denote by $\pi$ the mapping of symplectic fermions to superoperators. However, we aim to demonstrate that any representation of symplectic fermions cannot satisfy all the requirements for the dissipative theory. Thus, we intend to investigate not only a single representation, but an entire class of superoperator representations. Namely, we want to examine representation of the form $\pmb{\pi}_S(\cdot) = S\pmb{\pi}(\cdot) S^{-1}$, where $S$ is an arbitrary non-degenerate matrix. This approach is motivated by fermion algebra symmetry, and encompasses all possible symplectic fermion representations on ordinary fermion single-mode density matrices.

Conventional definition of $\pmb{\pi}$ is (the single mode parity operator has a simple form $\Pi = i \psi \bar\psi$):
\begin{align}
  \label{symplecticfermion-superrepresentation}
  \pmb{\pi}(J^-) \rho &= \psi\rho + \Pi \rho \Pi \psi = \psi \rho + \psi \bar\psi \rho \bar\psi, \\
  \pmb{\pi}(J^+) \rho &= \bar\psi \rho + \Pi \rho \Pi \bar\psi = \bar\psi \rho - \psi \bar\psi \rho \psi,
\end{align}
where $\rho$ is a fermion zero mode density matrix.

Using Pauli matrices for fermion $\psi, \, \bar\psi$ representation, we can write an explicit matrix for $\pmb{\pi}(\mathds{L}_0)$, where $\mathds{L}_0$ is one of the Virasoro generators. However, the matrix by itself is not particularly interesting. Its JNF is more important, because it is invariant under similarity transformation. As a result it remains the same for the all representation $\pmb{\pi}_S$:
\begin{equation}
\label{symplecticJNF}
 T_S \pmb{\pi}_S(\mathds{L}_0)T_S^{-1}= T\pmb{\pi}(\mathds{L}_0) T^{-1}= \begin{pmatrix}
    0 & 0 & 0 & 0 \\
    0 & 0 & 1 & 0 \\
    0 & 0 & 0 & 0 \\
    0 & 0 & 0 & 0
  \end{pmatrix},
\end{equation}
where $T_S $ and $T$ are appropriate transformation matrices.

Consequently, if we want to work with $\pmb{\pi}_S(\mathds{L}_0)$ directly, we need to look only at matrices with the JNF as in Eq.~\eqref{symplecticJNF}. On the other hand, the matrices $A$ with this JNF can be expressed as $ A_{ij} = \xi_i \eta_j$, where $\sum \xi_i \eta_i = 0 $. Thus, we have eight parameters to describe all possible representations of $\mathds{L}_0$ (actually, only seven independent parameters, because one can simultaneously rescale $\xi \mapsto \lambda \xi$ and $\eta \mapsto \lambda^{-1}\eta$  without affecting the form of $A$).

Taking into account that we work with superoperators, it is more convenient to use a "supermatrix" basis indexed by 4 indices for the density matrices. In this basis, superoperator matrices have the following elements: $A = \sum A_{ij,kl} E_{ij,kl}$. It reflects the fact that a superoperator acts on density matrices with two indices. Therefore, we will write a rank-$1$ Liouvillian as
\begin{equation}
  \label{one-rank-liouvillian}
  \pmb{\pi}_S(\mathds{L}_0) = \xi_{ij}\eta_{kl},
\end{equation}
where both $\xi$ and $\eta$ depend on $S$.

Next, we want to express the requirements of the hermiticity preservation, the trace preservation, and the complete positivity in terms of the parameters $\xi $ and $\eta$. 

Firstly, we consider how to treat the hermiticity preservation requirement. For an arbitrary density matrix $ \rho $ its Hermitian conjugation can be trivially written as $\rho^{\dagger} = \pmb{C}\pmb{P} \rho$, where $\pmb{C}$ is the complex conjugation operator, and $\pmb{P}$ is the transposition operator $ (\pmb{P}\rho)_{ij} = \rho_{ji}$.

Therefore, for $\pmb{\pi}_S(\mathds{L}_0)$ to preserve hermiticity, it needs to satisfy $\pmb{C} \pmb{P} \pmb{\pi}_S(\mathds{L}_0) \pmb{C}\pmb{P}  = \pmb{\pi}_S(\mathds{L}_0)$, or element-wise $\xi_{ji}^*\eta_{lk}^* = \xi_{ij} \eta_{kl}$.

Secondly, we want the time evolution generated by $\pmb{\pi}_S(\mathds{L}_0)$ to preserve the trace $ \tr\left[\exp\pmb{\pi}_S(\mathds{L}_0)t  \rho \right] = \tr\rho$. Using the nilpotency of the Liouvillian we obtain $ \exp\left[\pmb{\pi}_S(\mathds{L}_0)t\right] = {\mathds 1} + \pmb{\pi}_S(\mathds{L}_0) t$. Therefore, the superoperator~$\mathds{L}_0$ has to satisfy $ \tr\left[\pmb{\pi}_S(\mathds{L}_0) t\right] = 0$. In other words, the image of $\pmb{\pi}_S(\mathds{L}_0)$ needs to contain only traceless density matrices. But the image of $ \pmb{\pi}_S(\mathds{L}_0) $ is just a linear span of $\xi$. It means that $\xi$ needs to be traceless, or $\xi_{11} = - \xi_{22}$, which is another condition on parameters $\xi, \eta$.

Finally, we study the requirement of $\pmb{\pi}_S(\mathds{L}_0) $ being completely positive. To examine the complete positivity of $\pmb{\pi}_S(\mathds{L}_0)$ we use use Choi's theorem \cite{Choi1975}. It states that for a superoperator acting on $n\times n$-dimensional density matrices to be completely positive, the following $n^2 \times n^2$ dimensional Choi matrix $C$ has to be positive
\begin{equation}
\label{choi_matrix}
  C = \sum E_{ij} \otimes \mathcal{L}(E_{ij}),
\end{equation}
where $(E_{ij})_{pq} = \delta_{pi} \delta_{qj}$ is a density matrix basis.

We then plug the rank-$1$ decomposition of $\pmb{\pi}_S(\mathds{L}_0)$ from Eq.~\eqref{one-rank-liouvillian} into the Choi matrix~\eqref{choi_matrix}: $ C_{ij,kl} = \eta_{ik}\xi_{jl}$. Using Sylvester's criterion for non-negative matrices we are able to rewrite the requirement of $\pmb{\pi}_S(\mathds{L}_0)$ being completely positive as a system of inequalities on parameters $\xi, \, \eta$.

As solving the system of requirements on the parameters $\xi, \eta$ is rather straightforward, we do not describe it explicitly. It turns out that we cannot satisfy the three requirements simultaneously. Moreover, it appears that the trace preservation is inconsistent with the complete positivity.

Thus, it seems impossible to realize symplectic fermions as Liouvillian dynamics. 
Generally, the results of this appendix illustrate that not every non-unitary theory can be mapped onto Markovian dissipative dynamics. 
Further studies of logarithmic CFTs in the context of dissipative physical systems remains a subject for future research.

\bibliography{references}

\begin{thebibliography}{57}%
\makeatletter
\providecommand \@ifxundefined [1]{%
 \@ifx{#1\undefined}
}%
\providecommand \@ifnum [1]{%
 \ifnum #1\expandafter \@firstoftwo
 \else \expandafter \@secondoftwo
 \fi
}%
\providecommand \@ifx [1]{%
 \ifx #1\expandafter \@firstoftwo
 \else \expandafter \@secondoftwo
 \fi
}%
\providecommand \natexlab [1]{#1}%
\providecommand \enquote  [1]{``#1''}%
\providecommand \bibnamefont  [1]{#1}%
\providecommand \bibfnamefont [1]{#1}%
\providecommand \citenamefont [1]{#1}%
\providecommand \href@noop [0]{\@secondoftwo}%
\providecommand \href [0]{\begingroup \@sanitize@url \@href}%
\providecommand \@href[1]{\@@startlink{#1}\@@href}%
\providecommand \@@href[1]{\endgroup#1\@@endlink}%
\providecommand \@sanitize@url [0]{\catcode `\\12\catcode `\$12\catcode
  `\&12\catcode `\#12\catcode `\^12\catcode `\_12\catcode `\%12\relax}%
\providecommand \@@startlink[1]{}%
\providecommand \@@endlink[0]{}%
\providecommand \url  [0]{\begingroup\@sanitize@url \@url }%
\providecommand \@url [1]{\endgroup\@href {#1}{\urlprefix }}%
\providecommand \urlprefix  [0]{URL }%
\providecommand \Eprint [0]{\href }%
\providecommand \doibase [0]{https://doi.org/}%
\providecommand \selectlanguage [0]{\@gobble}%
\providecommand \bibinfo  [0]{\@secondoftwo}%
\providecommand \bibfield  [0]{\@secondoftwo}%
\providecommand \translation [1]{[#1]}%
\providecommand \BibitemOpen [0]{}%
\providecommand \bibitemStop [0]{}%
\providecommand \bibitemNoStop [0]{.\EOS\space}%
\providecommand \EOS [0]{\spacefactor3000\relax}%
\providecommand \BibitemShut  [1]{\csname bibitem#1\endcsname}%
\let\auto@bib@innerbib\@empty
\bibitem [{\citenamefont {Sachdev}(2011)}]{Sachdev2011}%
  \BibitemOpen
  \bibfield  {author} {\bibinfo {author} {\bibfnamefont {S.}~\bibnamefont
  {Sachdev}},\ }\href {https://doi.org/10.1017/CBO9780511973765} {\emph
  {\bibinfo {title} {Quantum Phase Transitions}}},\ \bibinfo {edition} {2nd}\
  ed.\ (\bibinfo  {publisher} {Cambridge University Press},\ \bibinfo {year}
  {2011})\BibitemShut {NoStop}%
\bibitem [{\citenamefont {Di~Francesco}\ \emph {et~al.}(1997)\citenamefont
  {Di~Francesco}, \citenamefont {Mathieu},\ and\ \citenamefont
  {Senechal}}]{DiFrancesco1997}%
  \BibitemOpen
  \bibfield  {author} {\bibinfo {author} {\bibfnamefont {P.}~\bibnamefont
  {Di~Francesco}}, \bibinfo {author} {\bibfnamefont {P.}~\bibnamefont
  {Mathieu}},\ and\ \bibinfo {author} {\bibfnamefont {D.}~\bibnamefont
  {Senechal}},\ }\href {https://doi.org/10.1007/978-1-4612-2256-9} {\emph
  {\bibinfo {title} {{Conformal Field Theory}}}},\ Graduate Texts in
  Contemporary Physics\ (\bibinfo  {publisher} {Springer-Verlag},\ \bibinfo
  {address} {New York},\ \bibinfo {year} {1997})\BibitemShut {NoStop}%
\bibitem [{\citenamefont {Polchinski}(1988)}]{Polchinski1988}%
  \BibitemOpen
  \bibfield  {author} {\bibinfo {author} {\bibfnamefont {J.}~\bibnamefont
  {Polchinski}},\ }\bibfield  {title} {\bibinfo {title} {Scale and conformal
  invariance in quantum field theory},\ }\href
  {https://doi.org/10.1016/0550-3213(88)90179-4} {\bibfield  {journal}
  {\bibinfo  {journal} {Nucl. Phys. B}\ }\textbf {\bibinfo {volume} {303}},\
  \bibinfo {pages} {226} (\bibinfo {year} {1988})}\BibitemShut {NoStop}%
\bibitem [{\citenamefont {Nakayama}(2014)}]{Nakayama2014}%
  \BibitemOpen
  \bibfield  {author} {\bibinfo {author} {\bibfnamefont {Y.}~\bibnamefont
  {Nakayama}},\ }\href@noop {} {\bibinfo {title} {Scale invariance vs conformal
  invariance}} (\bibinfo {year} {2014}),\ \Eprint
  {https://arxiv.org/abs/1302.0884} {arXiv:1302.0884 [hep-th]} \BibitemShut
  {NoStop}%
\bibitem [{\citenamefont {Vojta}(2003)}]{Vojta2003}%
  \BibitemOpen
  \bibfield  {author} {\bibinfo {author} {\bibfnamefont {M.}~\bibnamefont
  {Vojta}},\ }\bibfield  {title} {\bibinfo {title} {Quantum phase
  transitions},\ }\href {https://doi.org/10.1088/0034-4885/66/12/r01}
  {\bibfield  {journal} {\bibinfo  {journal} {Rep. Prog. Phys.}\ }\textbf
  {\bibinfo {volume} {66}},\ \bibinfo {pages} {2069} (\bibinfo {year}
  {2003})}\BibitemShut {NoStop}%
\bibitem [{\citenamefont {Kardar}\ \emph {et~al.}(1986)\citenamefont {Kardar},
  \citenamefont {Parisi},\ and\ \citenamefont {Zhang}}]{Kardar1986}%
  \BibitemOpen
  \bibfield  {author} {\bibinfo {author} {\bibfnamefont {M.}~\bibnamefont
  {Kardar}}, \bibinfo {author} {\bibfnamefont {G.}~\bibnamefont {Parisi}},\
  and\ \bibinfo {author} {\bibfnamefont {Y.-C.}\ \bibnamefont {Zhang}},\
  }\bibfield  {title} {\bibinfo {title} {Dynamic scaling of growing
  interfaces},\ }\href {https://doi.org/10.1103/PhysRevLett.56.889} {\bibfield
  {journal} {\bibinfo  {journal} {Phys. Rev. Lett.}\ }\textbf {\bibinfo
  {volume} {56}},\ \bibinfo {pages} {889} (\bibinfo {year} {1986})}\BibitemShut
  {NoStop}%
\bibitem [{\citenamefont {Krug}(1997)}]{Krug1997}%
  \BibitemOpen
  \bibfield  {author} {\bibinfo {author} {\bibfnamefont {J.}~\bibnamefont
  {Krug}},\ }\bibfield  {title} {\bibinfo {title} {Origins of scale invariance
  in growth processes},\ }\href {https://doi.org/10.1080/00018739700101498}
  {\bibfield  {journal} {\bibinfo  {journal} {Adv. Phys.}\ }\textbf {\bibinfo
  {volume} {46}},\ \bibinfo {pages} {139} (\bibinfo {year} {1997})}\BibitemShut
  {NoStop}%
\bibitem [{\citenamefont {Halpin-Healy}\ and\ \citenamefont
  {Zhang}(1995)}]{HalpinHealy1995}%
  \BibitemOpen
  \bibfield  {author} {\bibinfo {author} {\bibfnamefont {T.}~\bibnamefont
  {Halpin-Healy}}\ and\ \bibinfo {author} {\bibfnamefont {Y.-C.}\ \bibnamefont
  {Zhang}},\ }\bibfield  {title} {\bibinfo {title} {Kinetic roughening
  phenomena, stochastic growth, directed polymers and all that. aspects of
  multidisciplinary statistical mechanics},\ }\href
  {https://doi.org/10.1016/0370-1573(94)00087-j} {\bibfield  {journal}
  {\bibinfo  {journal} {Phys. Rep.}\ }\textbf {\bibinfo {volume} {254}},\
  \bibinfo {pages} {215} (\bibinfo {year} {1995})}\BibitemShut {NoStop}%
\bibitem [{\citenamefont {Aharonov}(2000)}]{Aharonov2000_PhysRevA.62.062311}%
  \BibitemOpen
  \bibfield  {author} {\bibinfo {author} {\bibfnamefont {D.}~\bibnamefont
  {Aharonov}},\ }\bibfield  {title} {\bibinfo {title} {Quantum to classical
  phase transition in noisy quantum computers},\ }\href
  {https://doi.org/10.1103/PhysRevA.62.062311} {\bibfield  {journal} {\bibinfo
  {journal} {Phys. Rev. A}\ }\textbf {\bibinfo {volume} {62}},\ \bibinfo
  {pages} {062311} (\bibinfo {year} {2000})}\BibitemShut {NoStop}%
\bibitem [{\citenamefont {Li}\ \emph {et~al.}(2019)\citenamefont {Li},
  \citenamefont {Chen},\ and\ \citenamefont
  {Fisher}}]{Fisher2019_PhysRevB.100.134306}%
  \BibitemOpen
  \bibfield  {author} {\bibinfo {author} {\bibfnamefont {Y.}~\bibnamefont
  {Li}}, \bibinfo {author} {\bibfnamefont {X.}~\bibnamefont {Chen}},\ and\
  \bibinfo {author} {\bibfnamefont {M.~P.~A.}\ \bibnamefont {Fisher}},\
  }\bibfield  {title} {\bibinfo {title} {Measurement-driven entanglement
  transition in hybrid quantum circuits},\ }\href
  {https://doi.org/10.1103/PhysRevB.100.134306} {\bibfield  {journal} {\bibinfo
   {journal} {Phys. Rev. B}\ }\textbf {\bibinfo {volume} {100}},\ \bibinfo
  {pages} {134306} (\bibinfo {year} {2019})}\BibitemShut {NoStop}%
\bibitem [{\citenamefont {Skinner}\ \emph {et~al.}(2019)\citenamefont
  {Skinner}, \citenamefont {Ruhman},\ and\ \citenamefont
  {Nahum}}]{Skinner2019_PhysRevX.9.031009}%
  \BibitemOpen
  \bibfield  {author} {\bibinfo {author} {\bibfnamefont {B.}~\bibnamefont
  {Skinner}}, \bibinfo {author} {\bibfnamefont {J.}~\bibnamefont {Ruhman}},\
  and\ \bibinfo {author} {\bibfnamefont {A.}~\bibnamefont {Nahum}},\ }\bibfield
   {title} {\bibinfo {title} {Measurement-induced phase transitions in the
  dynamics of entanglement},\ }\href
  {https://doi.org/10.1103/PhysRevX.9.031009} {\bibfield  {journal} {\bibinfo
  {journal} {Phys. Rev. X}\ }\textbf {\bibinfo {volume} {9}},\ \bibinfo {pages}
  {031009} (\bibinfo {year} {2019})}\BibitemShut {NoStop}%
\bibitem [{\citenamefont {Iaconis}\ \emph {et~al.}(2020)\citenamefont
  {Iaconis}, \citenamefont {Lucas},\ and\ \citenamefont {Chen}}]{Iaconis2020}%
  \BibitemOpen
  \bibfield  {author} {\bibinfo {author} {\bibfnamefont {J.}~\bibnamefont
  {Iaconis}}, \bibinfo {author} {\bibfnamefont {A.}~\bibnamefont {Lucas}},\
  and\ \bibinfo {author} {\bibfnamefont {X.}~\bibnamefont {Chen}},\ }\bibfield
  {title} {\bibinfo {title} {Measurement-induced phase transitions in quantum
  automaton circuits},\ }\href {https://doi.org/10.1103/PhysRevB.102.224311}
  {\bibfield  {journal} {\bibinfo  {journal} {Phys. Rev. B}\ }\textbf {\bibinfo
  {volume} {102}},\ \bibinfo {pages} {224311} (\bibinfo {year}
  {2020})}\BibitemShut {NoStop}%
\bibitem [{\citenamefont {Gong}\ \emph {et~al.}(2018)\citenamefont {Gong},
  \citenamefont {Hamazaki},\ and\ \citenamefont {Ueda}}]{Gong2018}%
  \BibitemOpen
  \bibfield  {author} {\bibinfo {author} {\bibfnamefont {Z.}~\bibnamefont
  {Gong}}, \bibinfo {author} {\bibfnamefont {R.}~\bibnamefont {Hamazaki}},\
  and\ \bibinfo {author} {\bibfnamefont {M.}~\bibnamefont {Ueda}},\ }\bibfield
  {title} {\bibinfo {title} {Discrete time-crystalline order in cavity and
  circuit {QED} systems},\ }\href
  {https://doi.org/10.1103/PhysRevLett.120.040404} {\bibfield  {journal}
  {\bibinfo  {journal} {Phys. Rev. Lett.}\ }\textbf {\bibinfo {volume} {120}},\
  \bibinfo {pages} {040404} (\bibinfo {year} {2018})}\BibitemShut {NoStop}%
\bibitem [{\citenamefont {Bu\ifmmode~\check{c}\else \v{c}\fi{}a}\ and\
  \citenamefont {Jaksch}(2019)}]{Buca219}%
  \BibitemOpen
  \bibfield  {author} {\bibinfo {author} {\bibfnamefont {B.}~\bibnamefont
  {Bu\ifmmode~\check{c}\else \v{c}\fi{}a}}\ and\ \bibinfo {author}
  {\bibfnamefont {D.}~\bibnamefont {Jaksch}},\ }\bibfield  {title} {\bibinfo
  {title} {Dissipation induced nonstationarity in a quantum gas},\ }\href
  {https://doi.org/10.1103/PhysRevLett.123.260401} {\bibfield  {journal}
  {\bibinfo  {journal} {Phys. Rev. Lett.}\ }\textbf {\bibinfo {volume} {123}},\
  \bibinfo {pages} {260401} (\bibinfo {year} {2019})}\BibitemShut {NoStop}%
\bibitem [{\citenamefont {Muniz}\ \emph {et~al.}(2020)\citenamefont {Muniz},
  \citenamefont {Barberena}, \citenamefont {Lewis-Swan}, \citenamefont {Young},
  \citenamefont {Cline}, \citenamefont {Rey},\ and\ \citenamefont
  {Thompson}}]{Muniz2020}%
  \BibitemOpen
  \bibfield  {author} {\bibinfo {author} {\bibfnamefont {J.~A.}\ \bibnamefont
  {Muniz}}, \bibinfo {author} {\bibfnamefont {D.}~\bibnamefont {Barberena}},
  \bibinfo {author} {\bibfnamefont {R.~J.}\ \bibnamefont {Lewis-Swan}},
  \bibinfo {author} {\bibfnamefont {D.~J.}\ \bibnamefont {Young}}, \bibinfo
  {author} {\bibfnamefont {J.~R.~K.}\ \bibnamefont {Cline}}, \bibinfo {author}
  {\bibfnamefont {A.~M.}\ \bibnamefont {Rey}},\ and\ \bibinfo {author}
  {\bibfnamefont {J.~K.}\ \bibnamefont {Thompson}},\ }\bibfield  {title}
  {\bibinfo {title} {Exploring dynamical phase transitions with cold atoms
  in~an optical~ cavity},\ }\href {https://doi.org/10.1038/s41586-020-2224-x}
  {\bibfield  {journal} {\bibinfo  {journal} {Nature}\ }\textbf {\bibinfo
  {volume} {580}},\ \bibinfo {pages} {602} (\bibinfo {year}
  {2020})}\BibitemShut {NoStop}%
\bibitem [{\citenamefont {Roberts}\ and\ \citenamefont
  {Clerk}(2020)}]{Roberts2020}%
  \BibitemOpen
  \bibfield  {author} {\bibinfo {author} {\bibfnamefont {D.}~\bibnamefont
  {Roberts}}\ and\ \bibinfo {author} {\bibfnamefont {A.~A.}\ \bibnamefont
  {Clerk}},\ }\bibfield  {title} {\bibinfo {title} {Driven-dissipative quantum
  {Kerr} resonators: New exact solutions, photon blockade and quantum
  bistability},\ }\href {https://doi.org/10.1103/PhysRevX.10.021022} {\bibfield
   {journal} {\bibinfo  {journal} {Phys. Rev. X}\ }\textbf {\bibinfo {volume}
  {10}},\ \bibinfo {pages} {021022} (\bibinfo {year} {2020})}\BibitemShut
  {NoStop}%
\bibitem [{\citenamefont {Ke{\ss}ler}\ \emph {et~al.}(2020)\citenamefont
  {Ke{\ss}ler}, \citenamefont {Cosme}, \citenamefont {Georges}, \citenamefont
  {Mathey},\ and\ \citenamefont {Hemmerich}}]{Keler2020}%
  \BibitemOpen
  \bibfield  {author} {\bibinfo {author} {\bibfnamefont {H.}~\bibnamefont
  {Ke{\ss}ler}}, \bibinfo {author} {\bibfnamefont {J.~G.}\ \bibnamefont
  {Cosme}}, \bibinfo {author} {\bibfnamefont {C.}~\bibnamefont {Georges}},
  \bibinfo {author} {\bibfnamefont {L.}~\bibnamefont {Mathey}},\ and\ \bibinfo
  {author} {\bibfnamefont {A.}~\bibnamefont {Hemmerich}},\ }\bibfield  {title}
  {\bibinfo {title} {From a continuous to a discrete time crystal in a
  dissipative atom-cavity system},\ }\href
  {https://doi.org/10.1088/1367-2630/ab9fc0} {\bibfield  {journal} {\bibinfo
  {journal} {New J. Phys.}\ }\textbf {\bibinfo {volume} {22}},\ \bibinfo
  {pages} {085002} (\bibinfo {year} {2020})}\BibitemShut {NoStop}%
\bibitem [{\citenamefont {Ippoliti}\ \emph {et~al.}(2021)\citenamefont
  {Ippoliti}, \citenamefont {Kechedzhi}, \citenamefont {Moessner},
  \citenamefont {Sondhi},\ and\ \citenamefont {Khemani}}]{Ippoliti2021}%
  \BibitemOpen
  \bibfield  {author} {\bibinfo {author} {\bibfnamefont {M.}~\bibnamefont
  {Ippoliti}}, \bibinfo {author} {\bibfnamefont {K.}~\bibnamefont {Kechedzhi}},
  \bibinfo {author} {\bibfnamefont {R.}~\bibnamefont {Moessner}}, \bibinfo
  {author} {\bibfnamefont {S.~L.}\ \bibnamefont {Sondhi}},\ and\ \bibinfo
  {author} {\bibfnamefont {V.}~\bibnamefont {Khemani}},\ }\bibfield  {title}
  {\bibinfo {title} {Many-body physics in the {NISQ} era: Quantum programming a
  discrete time crystal},\ }\href {https://doi.org/10.1103/PRXQuantum.2.030346}
  {\bibfield  {journal} {\bibinfo  {journal} {PRX Quantum}\ }\textbf {\bibinfo
  {volume} {2}},\ \bibinfo {pages} {030346} (\bibinfo {year}
  {2021})}\BibitemShut {NoStop}%
\bibitem [{\citenamefont {Bakker}\ \emph {et~al.}(2022)\citenamefont {Bakker},
  \citenamefont {Bahovadinov}, \citenamefont {Kurlov}, \citenamefont {Gritsev},
  \citenamefont {Fedorov},\ and\ \citenamefont {Krimer}}]{Bakker2022}%
  \BibitemOpen
  \bibfield  {author} {\bibinfo {author} {\bibfnamefont {L.~R.}\ \bibnamefont
  {Bakker}}, \bibinfo {author} {\bibfnamefont {M.~S.}\ \bibnamefont
  {Bahovadinov}}, \bibinfo {author} {\bibfnamefont {D.~V.}\ \bibnamefont
  {Kurlov}}, \bibinfo {author} {\bibfnamefont {V.}~\bibnamefont {Gritsev}},
  \bibinfo {author} {\bibfnamefont {A.~K.}\ \bibnamefont {Fedorov}},\ and\
  \bibinfo {author} {\bibfnamefont {D.~O.}\ \bibnamefont {Krimer}},\ }\bibfield
   {title} {\bibinfo {title} {Driven-dissipative time crystalline phases in a
  two-mode bosonic system with kerr nonlinearity},\ }\href
  {https://doi.org/10.1103/PhysRevLett.129.250401} {\bibfield  {journal}
  {\bibinfo  {journal} {Phys. Rev. Lett.}\ }\textbf {\bibinfo {volume} {129}},\
  \bibinfo {pages} {250401} (\bibinfo {year} {2022})}\BibitemShut {NoStop}%
\bibitem [{\citenamefont {Ding}\ \emph {et~al.}(2020)\citenamefont {Ding},
  \citenamefont {Busche}, \citenamefont {Shi}, \citenamefont {Guo},\ and\
  \citenamefont {Adams}}]{Ding_PhysRevX.10.021023}%
  \BibitemOpen
  \bibfield  {author} {\bibinfo {author} {\bibfnamefont {D.-S.}\ \bibnamefont
  {Ding}}, \bibinfo {author} {\bibfnamefont {H.}~\bibnamefont {Busche}},
  \bibinfo {author} {\bibfnamefont {B.-S.}\ \bibnamefont {Shi}}, \bibinfo
  {author} {\bibfnamefont {G.-C.}\ \bibnamefont {Guo}},\ and\ \bibinfo {author}
  {\bibfnamefont {C.~S.}\ \bibnamefont {Adams}},\ }\bibfield  {title} {\bibinfo
  {title} {Phase diagram and self-organizing dynamics in a thermal ensemble of
  strongly interacting {Rydberg} atoms},\ }\href
  {https://doi.org/10.1103/PhysRevX.10.021023} {\bibfield  {journal} {\bibinfo
  {journal} {Phys. Rev. X}\ }\textbf {\bibinfo {volume} {10}},\ \bibinfo
  {pages} {021023} (\bibinfo {year} {2020})}\BibitemShut {NoStop}%
\bibitem [{\citenamefont {Ding}\ \emph {et~al.}(2022)\citenamefont {Ding},
  \citenamefont {Liu}, \citenamefont {Shi}, \citenamefont {Guo}, \citenamefont
  {M{\o}lmer},\ and\ \citenamefont {Adams}}]{Ding_2022}%
  \BibitemOpen
  \bibfield  {author} {\bibinfo {author} {\bibfnamefont {D.-S.}\ \bibnamefont
  {Ding}}, \bibinfo {author} {\bibfnamefont {Z.-K.}\ \bibnamefont {Liu}},
  \bibinfo {author} {\bibfnamefont {B.-S.}\ \bibnamefont {Shi}}, \bibinfo
  {author} {\bibfnamefont {G.-C.}\ \bibnamefont {Guo}}, \bibinfo {author}
  {\bibfnamefont {K.}~\bibnamefont {M{\o}lmer}},\ and\ \bibinfo {author}
  {\bibfnamefont {C.~S.}\ \bibnamefont {Adams}},\ }\bibfield  {title} {\bibinfo
  {title} {Enhanced metrology at the critical point of a many-body {Rydberg}
  atomic system},\ }\href {https://doi.org/10.1038/s41567-022-01777-8}
  {\bibfield  {journal} {\bibinfo  {journal} {Nature Phys.}\ }\textbf {\bibinfo
  {volume} {18}},\ \bibinfo {pages} {1447} (\bibinfo {year}
  {2022})}\BibitemShut {NoStop}%
\bibitem [{\citenamefont {Zanardi}\ \emph {et~al.}(2008)\citenamefont
  {Zanardi}, \citenamefont {Paris},\ and\ \citenamefont
  {Campos~Venuti}}]{Zanardi2008}%
  \BibitemOpen
  \bibfield  {author} {\bibinfo {author} {\bibfnamefont {P.}~\bibnamefont
  {Zanardi}}, \bibinfo {author} {\bibfnamefont {M.~G.~A.}\ \bibnamefont
  {Paris}},\ and\ \bibinfo {author} {\bibfnamefont {L.}~\bibnamefont
  {Campos~Venuti}},\ }\bibfield  {title} {\bibinfo {title} {Quantum criticality
  as a resource for quantum estimation},\ }\href
  {https://doi.org/10.1103/PhysRevA.78.042105} {\bibfield  {journal} {\bibinfo
  {journal} {Phys. Rev. A}\ }\textbf {\bibinfo {volume} {78}},\ \bibinfo
  {pages} {042105} (\bibinfo {year} {2008})}\BibitemShut {NoStop}%
\bibitem [{\citenamefont {Minganti}\ \emph {et~al.}(2018)\citenamefont
  {Minganti}, \citenamefont {Biella}, \citenamefont {Bartolo},\ and\
  \citenamefont {Ciuti}}]{Minganti2018}%
  \BibitemOpen
  \bibfield  {author} {\bibinfo {author} {\bibfnamefont {F.}~\bibnamefont
  {Minganti}}, \bibinfo {author} {\bibfnamefont {A.}~\bibnamefont {Biella}},
  \bibinfo {author} {\bibfnamefont {N.}~\bibnamefont {Bartolo}},\ and\ \bibinfo
  {author} {\bibfnamefont {C.}~\bibnamefont {Ciuti}},\ }\bibfield  {title}
  {\bibinfo {title} {Spectral theory of {Liouvillians} for dissipative phase
  transitions},\ }\href {https://doi.org/10.1103/PhysRevA.98.042118} {\bibfield
   {journal} {\bibinfo  {journal} {Phys. Rev. A}\ }\textbf {\bibinfo {volume}
  {98}},\ \bibinfo {pages} {042118} (\bibinfo {year} {2018})}\BibitemShut
  {NoStop}%
\bibitem [{\citenamefont {Henkel}\ and\ \citenamefont
  {Pleimling}(2010)}]{Henkel2010}%
  \BibitemOpen
  \bibfield  {author} {\bibinfo {author} {\bibfnamefont {M.}~\bibnamefont
  {Henkel}}\ and\ \bibinfo {author} {\bibfnamefont {M.}~\bibnamefont
  {Pleimling}},\ }\href {https://doi.org/10.1007/978-90-481-2869-3} {\emph
  {\bibinfo {title} {Non-Equilibrium Phase Transitions}}}\ (\bibinfo
  {publisher} {Springer Netherlands},\ \bibinfo {year} {2010})\BibitemShut
  {NoStop}%
\bibitem [{\citenamefont {Heyl}(2018)}]{Heyl2018}%
  \BibitemOpen
  \bibfield  {author} {\bibinfo {author} {\bibfnamefont {M.}~\bibnamefont
  {Heyl}},\ }\bibfield  {title} {\bibinfo {title} {Dynamical quantum phase
  transitions: a review},\ }\href {https://doi.org/10.1088/1361-6633/aaaf9a}
  {\bibfield  {journal} {\bibinfo  {journal} {Rep. Prog. Phys.}\ }\textbf
  {\bibinfo {volume} {81}},\ \bibinfo {pages} {054001} (\bibinfo {year}
  {2018})}\BibitemShut {NoStop}%
\bibitem [{\citenamefont {Cubitt}\ \emph {et~al.}(2015)\citenamefont {Cubitt},
  \citenamefont {Lucia}, \citenamefont {Michalakis},\ and\ \citenamefont
  {Perez-Garcia}}]{Cubitt2015}%
  \BibitemOpen
  \bibfield  {author} {\bibinfo {author} {\bibfnamefont {T.~S.}\ \bibnamefont
  {Cubitt}}, \bibinfo {author} {\bibfnamefont {A.}~\bibnamefont {Lucia}},
  \bibinfo {author} {\bibfnamefont {S.}~\bibnamefont {Michalakis}},\ and\
  \bibinfo {author} {\bibfnamefont {D.}~\bibnamefont {Perez-Garcia}},\
  }\bibfield  {title} {\bibinfo {title} {Stability of local quantum dissipative
  systems},\ }\href {https://doi.org/10.1007/s00220-015-2355-3} {\bibfield
  {journal} {\bibinfo  {journal} {Comm. Math. Phys.}\ }\textbf {\bibinfo
  {volume} {337}},\ \bibinfo {pages} {1275} (\bibinfo {year}
  {2015})}\BibitemShut {NoStop}%
\bibitem [{\citenamefont {Cardy}(2013)}]{Cardy2013}%
  \BibitemOpen
  \bibfield  {author} {\bibinfo {author} {\bibfnamefont {J.}~\bibnamefont
  {Cardy}},\ }\bibfield  {title} {\bibinfo {title} {Logarithmic conformal field
  theories as limits of ordinary {CFTs} and some physical applications},\
  }\href {https://doi.org/10.1088/1751-8113/46/49/494001} {\bibfield  {journal}
  {\bibinfo  {journal} {J. Phys. A: Math. Theor.}\ }\textbf {\bibinfo {volume}
  {46}},\ \bibinfo {pages} {494001} (\bibinfo {year} {2013})}\BibitemShut
  {NoStop}%
\bibitem [{\citenamefont {Creutzig}\ and\ \citenamefont
  {Ridout}(2013)}]{Creutzig2013}%
  \BibitemOpen
  \bibfield  {author} {\bibinfo {author} {\bibfnamefont {T.}~\bibnamefont
  {Creutzig}}\ and\ \bibinfo {author} {\bibfnamefont {D.}~\bibnamefont
  {Ridout}},\ }\bibfield  {title} {\bibinfo {title} {Logarithmic conformal
  field theory: beyond an introduction},\ }\href
  {https://doi.org/10.1088/1751-8113/46/49/494006} {\bibfield  {journal}
  {\bibinfo  {journal} {J. Phys. A: Math. Theor.}\ }\textbf {\bibinfo {volume}
  {46}},\ \bibinfo {pages} {494006} (\bibinfo {year} {2013})}\BibitemShut
  {NoStop}%
\bibitem [{\citenamefont {Berges}\ \emph {et~al.}(2008)\citenamefont {Berges},
  \citenamefont {Rothkopf},\ and\ \citenamefont {Schmidt}}]{Berges2008}%
  \BibitemOpen
  \bibfield  {author} {\bibinfo {author} {\bibfnamefont {J.}~\bibnamefont
  {Berges}}, \bibinfo {author} {\bibfnamefont {A.}~\bibnamefont {Rothkopf}},\
  and\ \bibinfo {author} {\bibfnamefont {J.}~\bibnamefont {Schmidt}},\
  }\bibfield  {title} {\bibinfo {title} {Nonthermal fixed points: Effective
  weak coupling for strongly correlated systems far from equilibrium},\ }\href
  {https://doi.org/10.1103/PhysRevLett.101.041603} {\bibfield  {journal}
  {\bibinfo  {journal} {Phys. Rev. Lett.}\ }\textbf {\bibinfo {volume} {101}},\
  \bibinfo {pages} {041603} (\bibinfo {year} {2008})}\BibitemShut {NoStop}%
\bibitem [{\citenamefont {Berges}\ and\ \citenamefont
  {Hoffmeister}(2009)}]{Berges2009}%
  \BibitemOpen
  \bibfield  {author} {\bibinfo {author} {\bibfnamefont {J.}~\bibnamefont
  {Berges}}\ and\ \bibinfo {author} {\bibfnamefont {G.}~\bibnamefont
  {Hoffmeister}},\ }\bibfield  {title} {\bibinfo {title} {Nonthermal fixed
  points and the functional renormalization group},\ }\href
  {https://doi.org/10.1016/j.nuclphysb.2008.12.017} {\bibfield  {journal}
  {\bibinfo  {journal} {Nucl. Phys. B}\ }\textbf {\bibinfo {volume} {813}},\
  \bibinfo {pages} {383} (\bibinfo {year} {2009})}\BibitemShut {NoStop}%
\bibitem [{\citenamefont {Berges}\ and\ \citenamefont
  {Mesterh{\'{a}}zy}(2012)}]{Berges2012}%
  \BibitemOpen
  \bibfield  {author} {\bibinfo {author} {\bibfnamefont {J.}~\bibnamefont
  {Berges}}\ and\ \bibinfo {author} {\bibfnamefont {D.}~\bibnamefont
  {Mesterh{\'{a}}zy}},\ }\bibfield  {title} {\bibinfo {title} {Introduction to
  the nonequilibrium functional renormalization group},\ }\href
  {https://doi.org/10.1016/j.nuclphysbps.2012.06.003} {\bibfield  {journal}
  {\bibinfo  {journal} {Nucl. Phys. B Proc. Suppl.}\ }\textbf {\bibinfo
  {volume} {228}},\ \bibinfo {pages} {37} (\bibinfo {year} {2012})}\BibitemShut
  {NoStop}%
\bibitem [{\citenamefont {Nowak}\ \emph {et~al.}(2012)\citenamefont {Nowak},
  \citenamefont {Schole}, \citenamefont {Sexty},\ and\ \citenamefont
  {Gasenzer}}]{Nowak2012}%
  \BibitemOpen
  \bibfield  {author} {\bibinfo {author} {\bibfnamefont {B.}~\bibnamefont
  {Nowak}}, \bibinfo {author} {\bibfnamefont {J.}~\bibnamefont {Schole}},
  \bibinfo {author} {\bibfnamefont {D.}~\bibnamefont {Sexty}},\ and\ \bibinfo
  {author} {\bibfnamefont {T.}~\bibnamefont {Gasenzer}},\ }\bibfield  {title}
  {\bibinfo {title} {Nonthermal fixed points, vortex statistics, and superfluid
  turbulence in an ultracold {Bose} gas},\ }\href
  {https://doi.org/10.1103/PhysRevA.85.043627} {\bibfield  {journal} {\bibinfo
  {journal} {Phys. Rev. A}\ }\textbf {\bibinfo {volume} {85}},\ \bibinfo
  {pages} {043627} (\bibinfo {year} {2012})}\BibitemShut {NoStop}%
\bibitem [{\citenamefont {Schole}\ \emph {et~al.}(2012)\citenamefont {Schole},
  \citenamefont {Nowak},\ and\ \citenamefont {Gasenzer}}]{Schole2012}%
  \BibitemOpen
  \bibfield  {author} {\bibinfo {author} {\bibfnamefont {J.}~\bibnamefont
  {Schole}}, \bibinfo {author} {\bibfnamefont {B.}~\bibnamefont {Nowak}},\ and\
  \bibinfo {author} {\bibfnamefont {T.}~\bibnamefont {Gasenzer}},\ }\bibfield
  {title} {\bibinfo {title} {Critical dynamics of a two-dimensional superfluid
  near a nonthermal fixed point},\ }\href
  {https://doi.org/10.1103/PhysRevA.86.013624} {\bibfield  {journal} {\bibinfo
  {journal} {Phys. Rev. A}\ }\textbf {\bibinfo {volume} {86}},\ \bibinfo
  {pages} {013624} (\bibinfo {year} {2012})}\BibitemShut {NoStop}%
\bibitem [{\citenamefont {Karl}\ \emph {et~al.}(2013)\citenamefont {Karl},
  \citenamefont {Nowak},\ and\ \citenamefont {Gasenzer}}]{Karl2013}%
  \BibitemOpen
  \bibfield  {author} {\bibinfo {author} {\bibfnamefont {M.}~\bibnamefont
  {Karl}}, \bibinfo {author} {\bibfnamefont {B.}~\bibnamefont {Nowak}},\ and\
  \bibinfo {author} {\bibfnamefont {T.}~\bibnamefont {Gasenzer}},\ }\bibfield
  {title} {\bibinfo {title} {Universal scaling at nonthermal fixed points of a
  two-component {Bose} gas},\ }\href
  {https://doi.org/10.1103/PhysRevA.88.063615} {\bibfield  {journal} {\bibinfo
  {journal} {Phys. Rev. A}\ }\textbf {\bibinfo {volume} {88}},\ \bibinfo
  {pages} {063615} (\bibinfo {year} {2013})}\BibitemShut {NoStop}%
\bibitem [{\citenamefont {Cristofano}\ \emph {et~al.}(2004)\citenamefont
  {Cristofano}, \citenamefont {Marotta},\ and\ \citenamefont
  {Naddeo}}]{Cristofano2004}%
  \BibitemOpen
  \bibfield  {author} {\bibinfo {author} {\bibfnamefont {G.}~\bibnamefont
  {Cristofano}}, \bibinfo {author} {\bibfnamefont {V.}~\bibnamefont
  {Marotta}},\ and\ \bibinfo {author} {\bibfnamefont {A.}~\bibnamefont
  {Naddeo}},\ }\bibfield  {title} {\bibinfo {title} {A twisted conformal field
  theory description of dissipative quantum mechanics},\ }\href
  {https://doi.org/https://doi.org/10.1016/j.nuclphysb.2003.12.013} {\bibfield
  {journal} {\bibinfo  {journal} {Nucl. Phys. B}\ }\textbf {\bibinfo {volume}
  {679}},\ \bibinfo {pages} {621} (\bibinfo {year} {2004})}\BibitemShut
  {NoStop}%
\bibitem [{\citenamefont {Nakamura}(2012)}]{Nakamura2012}%
  \BibitemOpen
  \bibfield  {author} {\bibinfo {author} {\bibfnamefont {S.}~\bibnamefont
  {Nakamura}},\ }\bibfield  {title} {\bibinfo {title} {Nonequilibrium phase
  transitions and a nonequilibrium critical point from {Anti--de Sitter} space
  and conformal field theory correspondence},\ }\href
  {https://doi.org/10.1103/PhysRevLett.109.120602} {\bibfield  {journal}
  {\bibinfo  {journal} {Phys. Rev. Lett.}\ }\textbf {\bibinfo {volume} {109}},\
  \bibinfo {pages} {120602} (\bibinfo {year} {2012})}\BibitemShut {NoStop}%
\bibitem [{\citenamefont {Dutta}\ \emph {et~al.}(2015)\citenamefont {Dutta},
  \citenamefont {Aeppli}, \citenamefont {Chakrabarti}, \citenamefont
  {Divakaran}, \citenamefont {Rosenbaum},\ and\ \citenamefont
  {Sen}}]{Dutta2015}%
  \BibitemOpen
  \bibfield  {author} {\bibinfo {author} {\bibfnamefont {A.}~\bibnamefont
  {Dutta}}, \bibinfo {author} {\bibfnamefont {G.}~\bibnamefont {Aeppli}},
  \bibinfo {author} {\bibfnamefont {B.~K.}\ \bibnamefont {Chakrabarti}},
  \bibinfo {author} {\bibfnamefont {U.}~\bibnamefont {Divakaran}}, \bibinfo
  {author} {\bibfnamefont {T.~F.}\ \bibnamefont {Rosenbaum}},\ and\ \bibinfo
  {author} {\bibfnamefont {D.}~\bibnamefont {Sen}},\ }\href
  {https://doi.org/10.1017/cbo9781107706057} {\emph {\bibinfo {title} {Quantum
  Phase Transitions in Transverse Field Spin Models}}}\ (\bibinfo  {publisher}
  {Cambridge University Press},\ \bibinfo {year} {2015})\BibitemShut {NoStop}%
\bibitem [{\citenamefont {Chang}\ \emph {et~al.}(2020)\citenamefont {Chang},
  \citenamefont {You}, \citenamefont {Wen},\ and\ \citenamefont
  {Ryu}}]{Chang2020}%
  \BibitemOpen
  \bibfield  {author} {\bibinfo {author} {\bibfnamefont {P.-Y.}\ \bibnamefont
  {Chang}}, \bibinfo {author} {\bibfnamefont {J.-S.}\ \bibnamefont {You}},
  \bibinfo {author} {\bibfnamefont {X.}~\bibnamefont {Wen}},\ and\ \bibinfo
  {author} {\bibfnamefont {S.}~\bibnamefont {Ryu}},\ }\bibfield  {title}
  {\bibinfo {title} {Entanglement spectrum and entropy in topological
  non-hermitian systems and nonunitary conformal field theory},\ }\href
  {https://doi.org/10.1103/PhysRevResearch.2.033069} {\bibfield  {journal}
  {\bibinfo  {journal} {Phys. Rev. Res.}\ }\textbf {\bibinfo {volume} {2}},\
  \bibinfo {pages} {033069} (\bibinfo {year} {2020})}\BibitemShut {NoStop}%
\bibitem [{\citenamefont {Lindblad}(1976)}]{Lindblad1976}%
  \BibitemOpen
  \bibfield  {author} {\bibinfo {author} {\bibfnamefont {G.}~\bibnamefont
  {Lindblad}},\ }\bibfield  {title} {\bibinfo {title} {On the generators of
  quantum dynamical semigroups},\ }\href {https://doi.org/10.1007/bf01608499}
  {\bibfield  {journal} {\bibinfo  {journal} {Comm. Math. Phys.}\ }\textbf
  {\bibinfo {volume} {48}},\ \bibinfo {pages} {119} (\bibinfo {year}
  {1976})}\BibitemShut {NoStop}%
\bibitem [{\citenamefont {Gorini}(1976)}]{Gorini1976}%
  \BibitemOpen
  \bibfield  {author} {\bibinfo {author} {\bibfnamefont {V.}~\bibnamefont
  {Gorini}},\ }\bibfield  {title} {\bibinfo {title} {Completely positive
  dynamical semigroups of {N}-level systems},\ }\href
  {https://doi.org/10.1063/1.522979} {\bibfield  {journal} {\bibinfo  {journal}
  {J. Math. Phys.}\ }\textbf {\bibinfo {volume} {17}},\ \bibinfo {pages} {821}
  (\bibinfo {year} {1976})}\BibitemShut {NoStop}%
\bibitem [{\citenamefont {Breuer}\ and\ \citenamefont
  {Petruccione}(2007)}]{Breuer2007}%
  \BibitemOpen
  \bibfield  {author} {\bibinfo {author} {\bibfnamefont {H.-P.}\ \bibnamefont
  {Breuer}}\ and\ \bibinfo {author} {\bibfnamefont {F.}~\bibnamefont
  {Petruccione}},\ }\href
  {https://doi.org/10.1093/acprof:oso/9780199213900.001.0001} {\emph {\bibinfo
  {title} {The Theory of Open Quantum Systems}}}\ (\bibinfo  {publisher}
  {Oxford University {PressOxford}},\ \bibinfo {year} {2007})\BibitemShut
  {NoStop}%
\bibitem [{\citenamefont {Rivas}\ and\ \citenamefont
  {Huelga}(2011)}]{Rivas2011}%
  \BibitemOpen
  \bibfield  {author} {\bibinfo {author} {\bibfnamefont {{\'A}.}~\bibnamefont
  {Rivas}}\ and\ \bibinfo {author} {\bibfnamefont {S.}~\bibnamefont {Huelga}},\
  }\href@noop {} {\emph {\bibinfo {title} {{Open quantum systems: an
  introduction}}}},\ SpringerBriefs in Physics\ (\bibinfo  {publisher}
  {Springer Berlin Heidelberg},\ \bibinfo {year} {2011})\BibitemShut {NoStop}%
\bibitem [{\citenamefont {Fisher}(1978)}]{Fisher1978}%
  \BibitemOpen
  \bibfield  {author} {\bibinfo {author} {\bibfnamefont {M.~E.}\ \bibnamefont
  {Fisher}},\ }\bibfield  {title} {\bibinfo {title} {{Yang-Lee} edge
  singularity and ${\ensuremath{\phi}}^{3}$ field theory},\ }\href
  {https://doi.org/10.1103/PhysRevLett.40.1610} {\bibfield  {journal} {\bibinfo
   {journal} {Phys. Rev. Lett.}\ }\textbf {\bibinfo {volume} {40}},\ \bibinfo
  {pages} {1610} (\bibinfo {year} {1978})}\BibitemShut {NoStop}%
\bibitem [{\citenamefont {Cardy}(1985)}]{Cardy1985}%
  \BibitemOpen
  \bibfield  {author} {\bibinfo {author} {\bibfnamefont {J.~L.}\ \bibnamefont
  {Cardy}},\ }\bibfield  {title} {\bibinfo {title} {Conformal invariance and
  the {Yang-Lee} edge singularity in two dimensions},\ }\href
  {https://doi.org/10.1103/PhysRevLett.54.1354} {\bibfield  {journal} {\bibinfo
   {journal} {Phys. Rev. Lett.}\ }\textbf {\bibinfo {volume} {54}},\ \bibinfo
  {pages} {1354} (\bibinfo {year} {1985})}\BibitemShut {NoStop}%
\bibitem [{\citenamefont {Xu}\ and\ \citenamefont
  {Zamolodchikov}(2022)}]{xu2022}%
  \BibitemOpen
  \bibfield  {author} {\bibinfo {author} {\bibfnamefont {H.-L.}\ \bibnamefont
  {Xu}}\ and\ \bibinfo {author} {\bibfnamefont {A.}~\bibnamefont
  {Zamolodchikov}},\ }\bibfield  {title} {\bibinfo {title} {{2D Ising} field
  theory in a magnetic field: the {Yang-Lee} singularity},\ }\href
  {https://doi.org/10.1007/jhep08(2022)057} {\bibfield  {journal} {\bibinfo
  {journal} {J. High Energy Phys.}\ }\textbf {\bibinfo {volume} {2022}}\bibinfo
   {number} { (8)}}\BibitemShut {NoStop}%
\bibitem [{\citenamefont {Bhaseen}\ \emph {et~al.}(2001)\citenamefont
  {Bhaseen}, \citenamefont {Caux}, \citenamefont {Kogan},\ and\ \citenamefont
  {Tsvelik}}]{BHASEEN2001465}%
  \BibitemOpen
\bibfield  {number} {  }\bibfield  {author} {\bibinfo {author} {\bibfnamefont
  {M.~J.}\ \bibnamefont {Bhaseen}}, \bibinfo {author} {\bibfnamefont {J.-S.}\
  \bibnamefont {Caux}}, \bibinfo {author} {\bibfnamefont {I.~I.}\ \bibnamefont
  {Kogan}},\ and\ \bibinfo {author} {\bibfnamefont {A.~M.}\ \bibnamefont
  {Tsvelik}},\ }\bibfield  {title} {\bibinfo {title} {Disordered dirac
  fermions: the marriage of three different approaches},\ }\href
  {https://doi.org/https://doi.org/10.1016/S0550-3213(01)00432-1} {\bibfield
  {journal} {\bibinfo  {journal} {Nucl. Phys. B}\ }\textbf {\bibinfo {volume}
  {618}},\ \bibinfo {pages} {465} (\bibinfo {year} {2001})}\BibitemShut
  {NoStop}%
\bibitem [{\citenamefont {Gorbenko}\ \emph {et~al.}(2018)\citenamefont
  {Gorbenko}, \citenamefont {Rychkov},\ and\ \citenamefont
  {Zan}}]{Gorbenko_2018}%
  \BibitemOpen
  \bibfield  {author} {\bibinfo {author} {\bibfnamefont {V.}~\bibnamefont
  {Gorbenko}}, \bibinfo {author} {\bibfnamefont {S.}~\bibnamefont {Rychkov}},\
  and\ \bibinfo {author} {\bibfnamefont {B.}~\bibnamefont {Zan}},\ }\bibfield
  {title} {\bibinfo {title} {Walking, weak first-order transitions, and complex
  {CFTs}},\ }\href {https://doi.org/10.1007/jhep10(2018)108} {\bibfield
  {journal} {\bibinfo  {journal} {J. High Energy Phys.}\ }\textbf {\bibinfo
  {volume} {2018}}\bibinfo  {number} { (10)}}\BibitemShut {NoStop}%
\bibitem [{\citenamefont {Castro-Alvaredo}\ \emph {et~al.}(2017)\citenamefont
  {Castro-Alvaredo}, \citenamefont {Doyon},\ and\ \citenamefont
  {Ravanini}}]{Castro-Alvaredo_2017}%
  \BibitemOpen
\bibfield  {number} {  }\bibfield  {author} {\bibinfo {author} {\bibfnamefont
  {O.~A.}\ \bibnamefont {Castro-Alvaredo}}, \bibinfo {author} {\bibfnamefont
  {B.}~\bibnamefont {Doyon}},\ and\ \bibinfo {author} {\bibfnamefont
  {F.}~\bibnamefont {Ravanini}},\ }\bibfield  {title} {\bibinfo {title}
  {Irreversibility of the renormalization group flow in non-unitary quantum
  field theory},\ }\href {https://doi.org/10.1088/1751-8121/aa8a10} {\bibfield
  {journal} {\bibinfo  {journal} {J. Phys. A: Math. Theor.}\ }\textbf {\bibinfo
  {volume} {50}},\ \bibinfo {pages} {424002} (\bibinfo {year}
  {2017})}\BibitemShut {NoStop}%
\bibitem [{\citenamefont {Prosen}(2008)}]{Prosen2008}%
  \BibitemOpen
  \bibfield  {author} {\bibinfo {author} {\bibfnamefont {T.}~\bibnamefont
  {Prosen}},\ }\bibfield  {title} {\bibinfo {title} {Third quantization: a
  general method to solve master equations for quadratic open {Fermi}
  systems},\ }\href {https://doi.org/10.1088/1367-2630/10/4/043026} {\bibfield
  {journal} {\bibinfo  {journal} {New J. Phys.}\ }\textbf {\bibinfo {volume}
  {10}},\ \bibinfo {pages} {043026} (\bibinfo {year} {2008})}\BibitemShut
  {NoStop}%
\bibitem [{\citenamefont {Kos}\ and\ \citenamefont {Prosen}(2017)}]{Kos2017}%
  \BibitemOpen
  \bibfield  {author} {\bibinfo {author} {\bibfnamefont {P.}~\bibnamefont
  {Kos}}\ and\ \bibinfo {author} {\bibfnamefont {T.}~\bibnamefont {Prosen}},\
  }\bibfield  {title} {\bibinfo {title} {Time-dependent correlation functions
  in open quadratic fermionic systems},\ }\href
  {https://doi.org/10.1088/1742-5468/aa9681} {\bibfield  {journal} {\bibinfo
  {journal} {J. Stat. Mech.}\ }\textbf {\bibinfo {volume} {2017}},\ \bibinfo
  {pages} {123103} (\bibinfo {year} {2017})}\BibitemShut {NoStop}%
\bibitem [{fre()}]{free_boson_H_note}%
  \BibitemOpen
  \href@noop {} {}\bibinfo {note} {We truncate the Hamiltonian to the
  right-moving modes only, left moving-modes could be included
  similarly.}\BibitemShut {Stop}%
\bibitem [{\citenamefont {Prosen}(2010)}]{Prosen2010}%
  \BibitemOpen
  \bibfield  {author} {\bibinfo {author} {\bibfnamefont {T.}~\bibnamefont
  {Prosen}},\ }\bibfield  {title} {\bibinfo {title} {Spectral theorem for the
  {Lindblad} equation for quadratic open fermionic systems},\ }\href
  {https://doi.org/10.1088/1742-5468/2010/07/p07020} {\bibfield  {journal}
  {\bibinfo  {journal} {J. Stat. Mech.}\ }\textbf {\bibinfo {volume} {2010}},\
  \bibinfo {pages} {P07020} (\bibinfo {year} {2010})}\BibitemShut {NoStop}%
\bibitem [{\citenamefont {Barthel}\ and\ \citenamefont
  {Zhang}(2022)}]{Barthel2022}%
  \BibitemOpen
  \bibfield  {author} {\bibinfo {author} {\bibfnamefont {T.}~\bibnamefont
  {Barthel}}\ and\ \bibinfo {author} {\bibfnamefont {Y.}~\bibnamefont
  {Zhang}},\ }\bibfield  {title} {\bibinfo {title} {Solving quasi-free and
  quadratic {Lindblad} master equations for open fermionic and bosonic
  systems},\ }\href {https://doi.org/10.1088/1742-5468/ac8e5c} {\bibfield
  {journal} {\bibinfo  {journal} {J. Stat. Mech.}\ }\textbf {\bibinfo {volume}
  {2022}},\ \bibinfo {pages} {113101} (\bibinfo {year} {2022})}\BibitemShut
  {NoStop}%
\bibitem [{\citenamefont {{B\'acsi}}\ \emph {et~al.}(2020)\citenamefont
  {{B\'acsi}}, \citenamefont {Moca},\ and\ \citenamefont
  {{D\'ora}}}]{Bacsi2020}%
  \BibitemOpen
  \bibfield  {author} {\bibinfo {author} {\bibfnamefont {{\'A}.}~\bibnamefont
  {{B\'acsi}}}, \bibinfo {author} {\bibfnamefont {C.~P.}\ \bibnamefont
  {Moca}},\ and\ \bibinfo {author} {\bibfnamefont {B.}~\bibnamefont
  {{D\'ora}}},\ }\bibfield  {title} {\bibinfo {title} {Dissipation-induced
  {Luttinger} liquid correlations in a one-dimensional fermi gas},\ }\href
  {https://doi.org/10.1103/PhysRevLett.124.136401} {\bibfield  {journal}
  {\bibinfo  {journal} {Phys. Rev. Lett.}\ }\textbf {\bibinfo {volume} {124}},\
  \bibinfo {pages} {136401} (\bibinfo {year} {2020})}\BibitemShut {NoStop}%
\bibitem [{\citenamefont {{Ferrara}}\ \emph {et~al.}(1973)\citenamefont
  {{Ferrara}}, \citenamefont {{Grillo}},\ and\ \citenamefont
  {{Gatto}}}]{Ferrara1973}%
  \BibitemOpen
  \bibfield  {author} {\bibinfo {author} {\bibfnamefont {S.}~\bibnamefont
  {{Ferrara}}}, \bibinfo {author} {\bibfnamefont {A.~F.}\ \bibnamefont
  {{Grillo}}},\ and\ \bibinfo {author} {\bibfnamefont {R.}~\bibnamefont
  {{Gatto}}},\ }\bibfield  {title} {\bibinfo {title} {{Tensor representations
  of conformal algebra and conformally covariant operator product expansion}},\
  }\href {https://doi.org/10.1016/0003-4916(73)90446-6} {\bibfield  {journal}
  {\bibinfo  {journal} {Ann. Phys.}\ }\textbf {\bibinfo {volume} {76}},\
  \bibinfo {pages} {161} (\bibinfo {year} {1973})}\BibitemShut {NoStop}%
\bibitem [{\citenamefont {{Polyakov}}(1974)}]{Polyakov1974}%
  \BibitemOpen
  \bibfield  {author} {\bibinfo {author} {\bibfnamefont {A.~M.}\ \bibnamefont
  {{Polyakov}}},\ }\bibfield  {title} {\bibinfo {title} {{Non-Hamiltonian
  approach to conformal quantum field theory}},\ }\href@noop {} {\bibfield
  {journal} {\bibinfo  {journal} {[Zh. Eksp. Teor. Fiz. {\bf 66}, 23 (1974)]
  Sov. Phys. JETP}\ }\textbf {\bibinfo {volume} {39}},\ \bibinfo {pages} {10}
  (\bibinfo {year} {1974})}\BibitemShut {NoStop}%
\bibitem [{\citenamefont {Choi}(1975)}]{Choi1975}%
  \BibitemOpen
  \bibfield  {author} {\bibinfo {author} {\bibfnamefont {M.-D.}\ \bibnamefont
  {Choi}},\ }\bibfield  {title} {\bibinfo {title} {Completely positive linear
  maps on complex matrices},\ }\href
  {https://doi.org/10.1016/0024-3795(75)90075-0} {\bibfield  {journal}
  {\bibinfo  {journal} {Lin. Alg. Appl.}\ }\textbf {\bibinfo {volume} {10}},\
  \bibinfo {pages} {285} (\bibinfo {year} {1975})}\BibitemShut {NoStop}%
\end{thebibliography}%
\end{document}